\documentclass[10pt]{JHEP3}
\JHEPspecialurl{http://jhep.sissa.it/JOURNAL/JHEP3.tar.gz}
\usepackage{bm}  
\usepackage{amsmath}     
\usepackage{epsfig,epsf}
\usepackage{graphicx}
\usepackage{epstopdf}
\newcommand{\beq}{\begin{equation}}
\newcommand{\eeq}{\end{equation}}
\def\eq#1{{(\ref{#1})}}
\def\fig#1{{Fig.~\ref{#1}}}

\newcommand{\be}{\begin{eqnarray}}
\newcommand{\ee}{\end{eqnarray}}

\newcommand{\Q}{\mathcal{Q}}
\newcommand{\Lb}{\left(}
\newcommand{\Rb}{\right)}
\newcommand{\as}{\alpha_s}

\newcommand{\Tr}{{\rm Tr}}

\newcommand{\un}{\underline}

\newcommand{\ben}{\begin{eqnarray*}}
\newcommand{\een}{\end{eqnarray*}}
\title{\LARGE \bf J/$\mathbf{\Psi}$  production in heavy ion collisions and gluon saturation}
\author{\large {Dmitri Kharzeev $^a$\,,\, Eugene Levin $^b$\,,\,Marzia Nardi $^c$\, and\,
Kirill Tuchin
$^{d,e}$} \vspace{3mm}\\
a) Department of Physics, Brookhaven National Laboratory,\\
Upton, New York 11973-5000, USA\vspace{3mm}\\ 
b) HEP Department, School of Physics,\\
Raymond and Beverly Sackler Faculty of Exact Science,\\
Tel Aviv University, Tel Aviv 69978, Israel\vspace{3mm}\\ 
c) Istituto   Nazionale  di Fisica   Nucleare, Sezione  di Torino, \\
via P.Giuria 1, I-10125 Torino, Italy \vspace{3mm}\\ 
d) Department of Physics and Astronomy,\\
Iowa State University, Ames, Iowa 50011, USA\vspace{3mm}\\
e) RIKEN BNL Research Center,\\
Upton, New York 11973-5000, USA\\
}

\preprint{BNL-NT-, TAUP-2883-08, RBRC-752\\
\today }

\abstract{We calculate the inclusive $ J/\psi$ production in
  heavy ion collisions including the effects of gluon saturation
in the wave functions of the colliding nuclei. We argue that the
dominant production mechanism  in proton--nucleus and
nucleus--nucleus collisions for heavy nuclei is different from the one in
hadron-hadron interactions. We find that the
rapidity distribution of primary $ J/\psi$ production is more peaked
around midrapidity than the analogous distribution in elementary $pp$ 
collisions. We discuss the consequences of this fact on the
experimentally observed  $ J/\psi$ suppression in $Au-Au$ collisions at
RHIC energies.}


\begin{document}

\section{Introduction}

Understanding the mechanism of $ J/\psi$ production has been a
challenge for over three decades. Despite a relatively large charm quark mass, the 
binding energy of $ J/\psi$ is quite small; therefore non-perturbative
corrections can be important. This is a likely cause of the difficulties encountered by 
perturbative QCD in 
describing the differential $ J/\psi$ production cross section and
its polarization. A significant effort has been invested into attempts
to uncover the mysteries of $ J/\psi$ production. Still, when confronted with the experimental
data the existing
approaches encounter problems that have to be cured by the introduction of additional 
adjustable parameters encoding the poorly understood dynamics (for a recent review see \cite{Brambilla:2004wf}).  

In this paper we develop a new approach to the $ J/\psi$ production in
nuclear reaction suggested by two of us in \cite{KT}. It was argued in
Ref.~\cite{KT} that at high energies
the dynamics of $ J/\psi$ production is determined mostly by the strength of the coherent quasi-classical fields of the nucleus.    
This approach yielded a reasonable description of the experimental
data, and in particular shone some light on the possible origin of  $x_F$
scaling in $p(d)A$ collisions observed in the data from
CERN\cite{Badier:1983dg}, FNAL\cite{Leitch:1999ea} and
RHIC\cite{Adler:2005ph}.  

In the context of high energy nuclear physics, it is important to
understand well the mechanism of $ J/\psi$ production since $ J/\psi$
suppression in heavy ion collisions could serve as a signal of the
Quark-Gluon Plasma \cite{Matsui:1986dk}.  
Motivated by the urgent 
necessity to understand the cold nuclear effects on  the $ J/\psi$
production in heavy ion  
collisions, we shall calculate the inclusive $ J/\psi$ production in
heavy ion collisions. In this paper we take into account only the cold
nuclear matter effects neglecting completely the dynamical effects
leading to the possible formation of the Quark Gluon Plasma.  

A systematic approach to the particle production in heavy ion
collisions at high energies has been developed in
Refs.~\cite{KLN,KLM,KL1,KN1} and will be referred to as the KLN
approach.  KLN assumes that the wave functions of the colliding nuclei can be described as the Color Glass Condensate
(CGC)  \cite{MV,JIMWLK} for which the most characteristic
property is the saturation of the parton density
\cite{GLR,MUQI,MV,Blaizot:nc}. These ideas have passed the first check
against the RHIC experimental data on multiplicities and rapidity
distributions (see Refs. \cite{KLN,KLM,KL1,KN1,KKT,Kovner:2003zj})
and, in this paper, we confront them with the rapidity distribution of  
 $J/\psi$ mesons at RHIC.   Our study in this paper is based on the
following  observations made   in
Ref. \cite{KT}: (1) The  mechanism of $ J/\psi$ production in
hadron-nucleus collisions  is different from the one in the
hadron-hadron interactions; (2) Inclusive  $J/\psi$ production and
inclusive $c\bar c$ production are dominated by different distance
scales.  

To explain our main idea, consider the $ J/\psi$ production in
hadron--hadron collisions. The leading contribution is given by the
two-gluon fusion 
\beq \label{I1}
G\,+\,G\,\to\, J/\psi\,+\,\mathrm{\,gluon}\,.
\eeq
This process is of the order $\mathcal{O}(\as^5)$: the partonic sub-process is of the order $\as^3$;  two additional powers of $\as$ arise from attaching the initial gluons to the colliding hadrons. The three-gluon fusion 
\beq \label{I2}
G\,+\,G\,+\,G\,\to\, J/\psi
\eeq
is parametrically suppressed as it is proportional to
$\mathcal{O}(\as^6)$. However, in hadron-nucleus collisions two of the initial
state gluons can be attached to the nucleus. This brings in an
additional enhancement by $A^{1/3}$. Since in the quasi-classical
approximation $\as^2A^{1/3}\sim 1$ we find that the three-gluon fusion
of \eq{I2} is actually \emph{enhanced} by $1/\as$ as compared to
\eq{I1}. Similar conclusion holds for heavy ion collisions.  

A particularly helpful insight into the nature of the contribution
\eq{I2} is obtained   if we note that three-gluon contribution \eq{I2}
is suppressed as compared to the two-gluon one \eq{I1} by an
additional factor $r^2$, where $(2m_c)^{-1}<r<(2m_c\as)^{-1}$.  This
factor arises since  we need to have three gluons in the area  of the
order of $ r^2$.  In other words, it means that this reaction 
originates from the higher-twist contribution. However,  in the hadron -
nucleus interactions the higher-twist contribution appears always in the
dimensionless  combination $r^2 Q_s^2$ with the saturation scale
$Q_s$. The saturation scale is proportional to $A^{1/3}$ which
compensates for the smallness of $r$.  
The dominance of the higher twist process \eq{I2} is main idea of \cite{KT} 
and we are going to develop it in this paper in the case of heavy ion collisions. 
Various aspects of multi-parton interactions generating the higher twist effects in
$ J/\psi$ production were considered previously in
\cite{Brodsky:1991dj,Kopeliovich:1984bf,
  Vogt:1991qd,Gavin:1991qk,Kharzeev:1993qd,Kharzeev:1995br,Clavelli:1985kg,Benesh:1994du,Fujii:2003ff,acp}.  
 
The paper is structured as follows: in Section 2 we describe the
production of $c\bar{c}$ pairs in proton-nucleus and nucleus-nucleus
collisions; in Sections 3 and 4 we calculate the inclusive $J/\psi$
production cross-section in these collisions. 
We present our numerical results in Section 5.

\section{Warm-up: inclusive production of $c \bar{c}$ pair with fixed relative momentum}
\subsection{Hadron-hadron collisions}


\FIGURE[h]{
\centerline{\epsfig{file=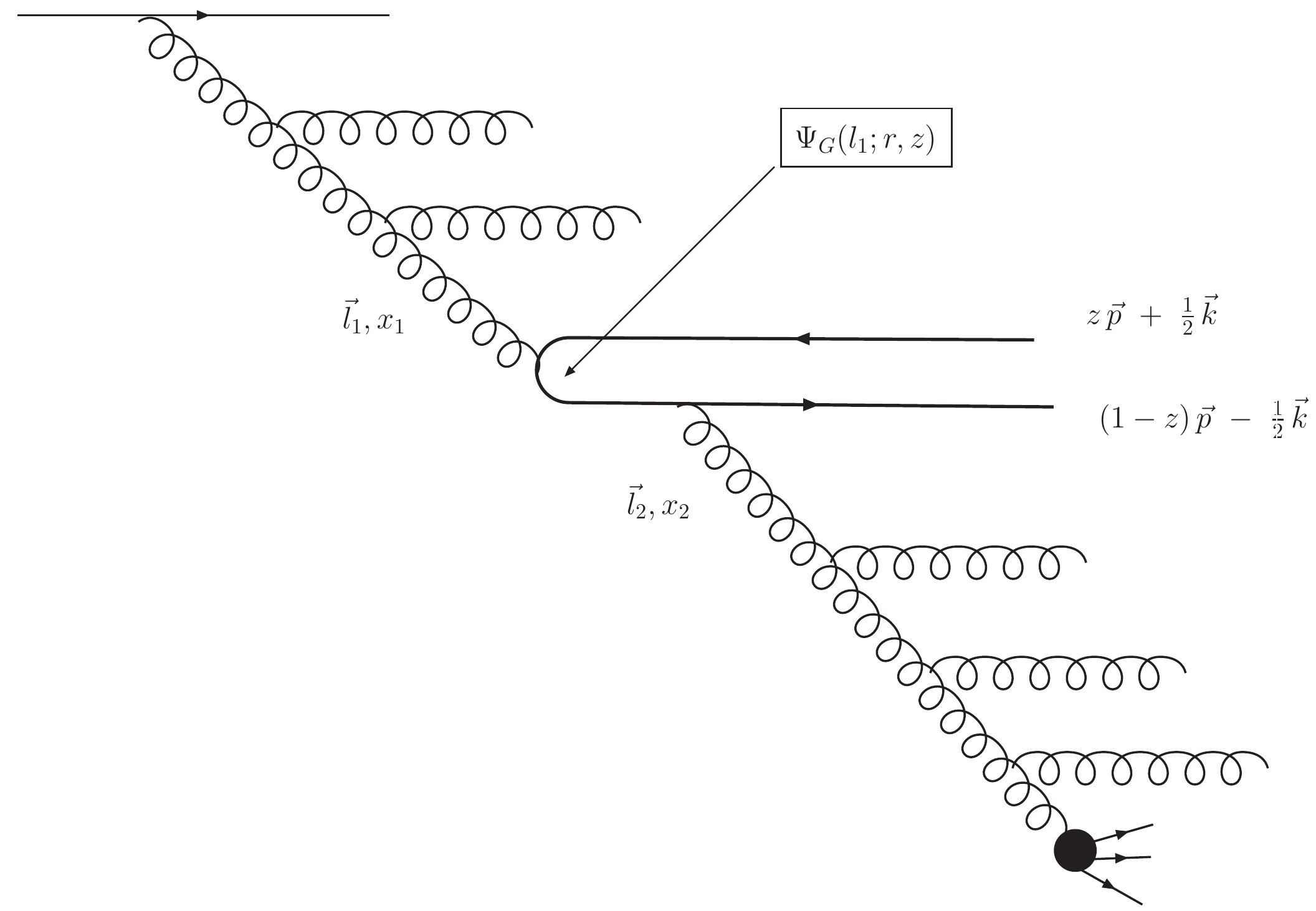,width=9cm}}
\caption{ The  process of inclusive $c \bar{c}$  production  with
  fixed relative momentum ($\vec{k}$)   in hadron-hadron  collision. 
\label{ccpp} }
}

The process of  $c \bar{c}$ production in a hadron-hadron collision is
shown in \fig{ccpp}. The corresponding cross section  integrated over the transverse momentum of
the pair, denoted by $\vec{p}\equiv \un p$ ($\un p^2\equiv p^2$) in \fig{ccpp},  is equal to the square
of the diagram in this figure and can be written as follows 
\begin{eqnarray}
\label{IE1}
\frac{ d \sigma(pp)}{d Y\, d^2k}&=&\frac{1}{(2\pi)^3}\frac{1}{2(N_c^2-1)}\frac{4\pi^2\as}{N_c} \sum_{s,s',\lambda}\int\frac{d^2 l_1}{\pi }\,\phi_G(x_1, \un l_1^2)
\,\int\,\frac{d^2  l_2}{2 \pi \un l^2_2 }\,\phi_G(x_2,\un l_2^2)\, \times \nonumber\\
&& 2\,\int d^2\,r \,d z\,\Psi_G(l_1,r,z)\Lb 1 - e^{i \un {l}_2
  \cdot \un{r}} \Rb\,e^{- i\frac{1}{2}\un{k}\cdot \un{r}}
\int d^2\,r' \,\Psi_G^*(l_1,r',z)\Lb 1 - e^{-i \un{l}_2 \cdot
  \un{r}'} \Rb\,e^{ i\frac{1}{2}\un{k}\cdot \un{r}'}
\end{eqnarray}
where $x_1\,=\,(m_{c,t}+m_{\bar c,t})\,e^{Y}/\sqrt{s}$ and
$x_2\,=\,(m_{c,t}+m_{\bar c,t})\,e^{-Y}/\sqrt{s}$~,
with $Y$ being the rapidity of quark and antiquark pair in the
center-of-mass frame,  $m^2_{c,t} \,=\,m^2_c + \un k_1^2$,
$m^2_{\bar c,t} \,=\,m^2_c + \un k_2^2$, $s,s'$ are the quark and anti-quark helicities and $\lambda$ is the gluon polarization.  
The function $\phi_G(x,\un l^2)$ is the probability to find
a gluon with given $x$ and transverse momentum $\un l$. It is related to the gluon distribution
function $xG(x,Q^2)$ as
\beq \label{IE2}
x G(x,Q^2)\,=\,\int^{Q^2}d l^2\,\phi(x,\un l^2)
\eeq
The factor 2 in front of \eq{IE1} is a consequence of the $s$-channel
unitarity by which the inelastic cross section equals
twice the imaginary part of the elastic scattering
amplitude. Eq.~\eq{ccpp} is written in the $k_T$-factorization
approach which is believed to be valid in hadron-hadron collisions at
not too high energies \cite{LRSS,CCH,CE}.

It is convenient to introduce the cross section for dipole--hadron
interaction in the form \cite{XS} 
\beq \label{IE3}
\sigma \Lb x, r^2 \Rb\,=\,\frac{8\pi^2\as}{N_c}\,\int \,\frac{d^2 l}{2
  \pi l^2}\,\Lb 1 - e^{ i\,\un{r} \cdot\,\un{l}}\Rb\,\phi\Lb x, \un l^2
\Rb\,.
\eeq
In the DGLAP approximation the dominant contribution to the integral over $\un l$ comes from the region $lr<2$ where it picks up the leading logarithmic contribution. Integrating first over all directions of the vector $\un l$ and then expanding the resulting Bessel function yields:
\beq
\sigma(x,r^2)\approx \frac{8\pi^2\as}{N_c}\int_0^{2/r} \frac{dl}{l}\,\frac{1}{4}r^2l^2\,\phi(x,l^2)
= \frac{\as\pi^2}{N_c}\,r^2\,xG(x,4/r^2)\,,
\eeq
where we used \eq{IE2}.

Using \eq{IE2} and \eq{IE3} it is easy to rewrite \eq{IE1} in the following form
\begin{eqnarray} \label{IE4}
\frac{d \sigma(pp)}{ d Y\, d^2k} &=& \frac{1}{(2\pi)^3}\frac{1}{2(N_c^2-1)}\sum_{s,s',\lambda} x_1 G(x_1, m^2_c)\, 
\times\nonumber\\
&&\int d^2 r\,\Psi_G(m_c, r, z=1/2)\,e^{-i\, \frac{1}{2}\un{r}\cdot \un{k} }\,
\int d^2 r'\,\Psi_G^*(m_c, r',z=1/2)\,e^{i\,\frac{1}{2} \un{r}' \cdot \un{k}}\,
\hat\sigma_{in}(x_2,r,r')
\end{eqnarray}
where 
\beq \label{XSIN}
\hat\sigma_{in}(x_2,r,r')\,\equiv\,
\sigma(x_2, r^2) + \sigma(x_2, r'^2)\,-\,\sigma(x_2,(\un{r} - \un{r}')^2)\,.
\eeq
(The $\hat\sigma$ notation is used to distinguish the \emph{dipole} cross section
defined in \eq{XSIN} from the inclusive heavy quark-antiquark
inelastic cross section we discuss later, see \eq{IE7}). In derivation
of \eq{IE4} we took into account only the  DGLAP contribution to $x_1
G\Lb x_1, m^2_c \Rb$ and we treated the $c$-quark as a
non-relativistic particle with $z = 1/2$. All these simplifications are
not important for our main results but allow for a more compact notations. 

The  gluon light-cone wave function is well-known
(Refs. \cite{dipole,WF,BRLE}). It has the simplest form for $z=1/2$
and $\un l^2_1/4 \,\ll\,m^2_c$, namely (see \eq{phiT}) 
\be
&&\Psi_G(m_c, r, z=1/2)\,=\,\frac{g \, t^a}{2\pi}\left[ i\, \frac{\un
  r\cdot \un \epsilon^\lambda}{r}\,m_c\,K_1 (r m_c)\,\lambda
\,\delta_{s,s'}\,+\,K_0( r m_c)\,s\,m_c(1+s\lambda) \delta_{s,-s'}
\right]  \,;\label{WF1}\\ 
&&\Phi_G(m_c,r,r',z=1/2)\,=\,\frac{1}{(2\pi)^3}\,
\frac{1}{2(N_c^2-1)}\sum_{\lambda,s,s'}\,\Psi_G(m_c, 
r, z=1/2)\Psi^*_G(m_c, r', z=1/2) 
\nonumber\\
 &&
\,=\, \frac{1}{(2\pi)^3}\,\frac{\as m^2_c}{\pi} \, \left[ \frac{1}{2}\frac{\un r
  \cdot \un r '}{r r'} 
K_1 \Lb r m_c\Rb K_1 \Lb r' m_c \Rb \,+\,K_0 \Lb r m_c\Rb K_0 \Lb
r' m_c \Rb \right]\,, \label{WF2} 
\ee
where $t^a$ is the Gell-Mann matrix and $\un
\epsilon^\lambda$ is the polarization vector. With these definition we can write \eq{IE4} as
\beq \label{last}
\frac{d \sigma(pp)}{ d Y\, d^2k} =  x_1 G(x_1, m^2_c)\, \int d^2 r
\int d^2 r'\,\Phi_G(m_c, r,r',z=1/2)\,e^{i\,\frac{1}{2} (\un{r}'-\un r) \cdot \un{k}}\,
\hat\sigma_{in}(x_2,r,r')
\eeq
In Appendix we give a detailed derivation of these formulas.

\subsection{Hadron--heavy nucleus collisions}

Production of  quark-antiquark
pairs in high energy proton-nucleus collisions and in DIS both in the
quasi-classical approximation of McLerran-Venugopalan model \cite{MV}
(summing powers of $\as^2 A^{1/3}$) and including quantum small-$x$
evolution (summing powers of $\as \ln\frac{1}{x} $) has been calculated in
Ref.~\cite{Tuchin:2004rb,Kovchegov:2006qn}. This process has been also
considered by other authors \cite{Gelis:2003vh,Blaizot:2004wv,KopTar}
who obtained similar, though less general, results. Phenomenological
applications have been addressed in details in
\cite{KhT,Tuchin:2007pf}. Using the results of
\cite{Tuchin:2004rb,Kovchegov:2006qn,Tuchin:2007pf} it is not
difficult to generalize the formulae of the previous subsection for the
case of $pA$ collisions. The details are given in Appendix. Here we
present a derivation that emphasizes the key physical issues.  
 
As one can see in \fig{ccpa} the quark-antiquark pair production in
hadron-nucleus interaction includes an additional elastic scattering
of dipoles with sizes $r$ and $r'$ as well as  inelastic interaction
at points $z_i$, which are the  longitudinal coordinates of nucleons in the nucleus \footnote{Note that
  $z$ ($z'$) appearing in \eq{WF1}, \eq{WF2} etc.\ denote the fraction
  of the gluon's light-cone momentum carried by the $c$-quark in the
  (complex conjugated) amplitude. $z_i$'s with $i=0,1,2,\ldots$ in
  \fig{ccpa} etc.\ denote the longitudinal coordinates of nucleons in
  the nucleus. These are two completely unrelated variables.}. 
\FIGURE[ht]{
\centerline{\epsfig{file=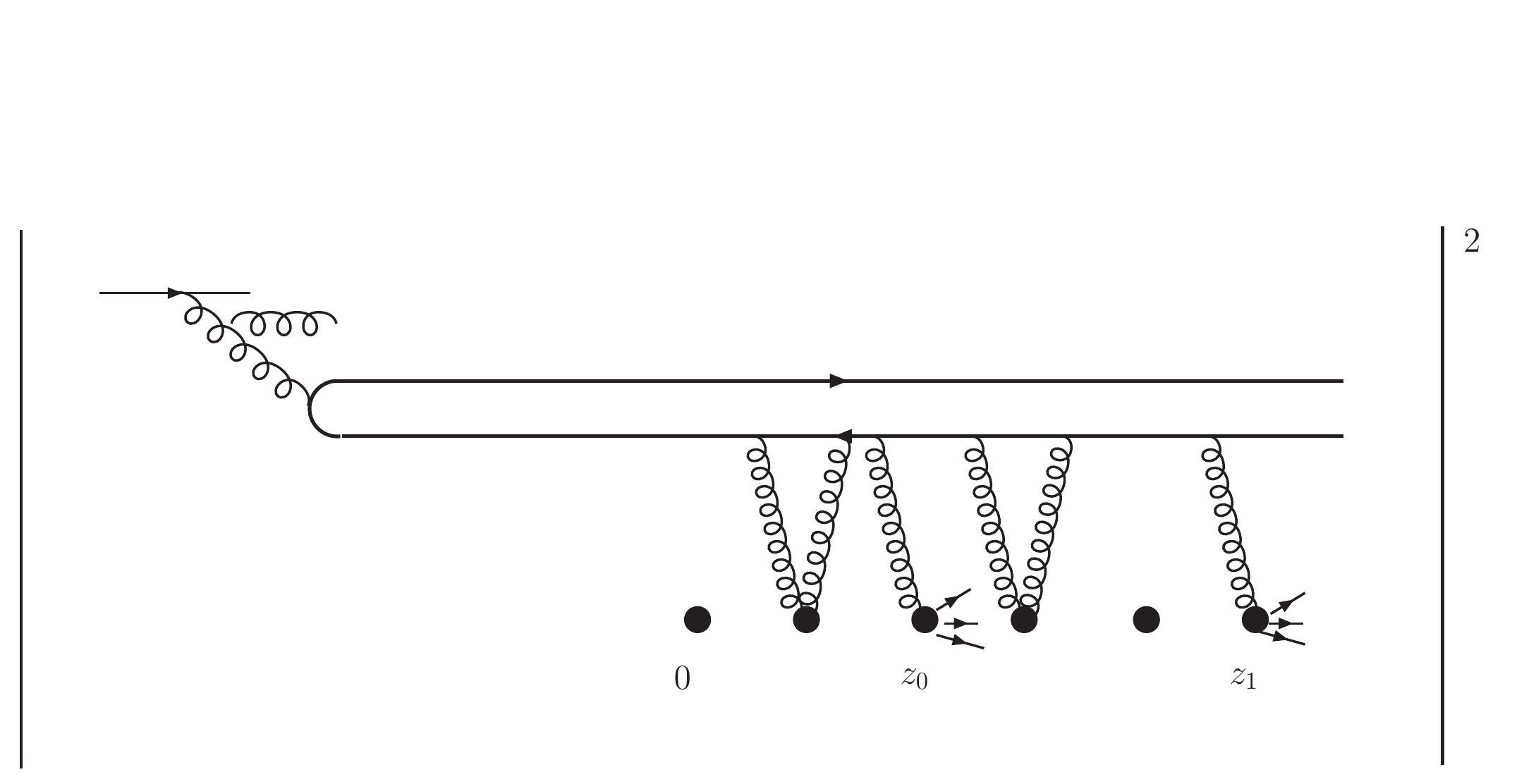,width=12cm}}
\caption{ The  process of inclusive $c \bar{c}$  production  with
  fixed relative momentum ($\un{k}$)   in hadron-nucleus
  collision. } 
\label{ccpa} }
To include  both processes we need to modify  \eq{IE1} in the following way
\be\label{IE5}
\frac{ d \sigma_{in}(pA)}{d Y \, d^2k\, d^2b}&=&
x_1 G(x_1, m^2_c)\, \int d^2 r
\int d^2 r'\,\Phi_G(m_c, r,r',z=1/2)\,e^{i\frac{1}{2} (\un{r}'-\un r) \cdot \un{k}}\nonumber\\ 
&& \times   \int^{2R_A}_{0}\rho\, \hat\sigma_{in}(x_2,r,r')\,d
z_0\,e^{-[\sigma(x_2,r^2) +\sigma(x_2,r'^2)]\,\rho\,2\,R_A }\nonumber\\
&&\times
\,\sum^{\infty}_{n=0} 
\int^{2 R_A}_{z_0}d\,z_{1}\, \dots \int^{2 R_A}_{z_{n -2}}
dz_{n-1}\int^{2 R_A}_{z_{n-1} }d z_n\,
\rho^n\,\hat\sigma^n_{in}(x_2,r,r')\, 
\ee
where $\rho$ is the density of the nucleons in  a  nucleus and $R_A$
is the nucleus radius. For brevity  we wrote \eq{IE5} for a
cylindrical nucleus.  In Sec.~\ref{sec:numerics}  we perform numerical analyses with  realistic nuclear density distributions. 

The factor $\exp\left\{-[\sigma(x_2,r^2) +\sigma(x_2,r'^2)]\rho\, 2R_A
\right\}$ in \eq{IE5} describes the fact that neither dipole with the
size $r$ (in the amplitude) nor dipole with the size $r'$ (in the
complex conjugate amplitude) interacts inelastically with the nucleons
of the nucleus between the points 0 and $z_1$ as well as between any
other pair of points $z_{i}$  and $z_{i-1}$, where $i=0,1,\ldots,
n-1$.  
 In deriving Eq.~\eq{IE5} we assumed that the collision energy is high
 enough so that the quark-antiquark pair is produced  
long before it starts to interact with the nucleus. This corresponds
to coherent interaction of the pair with all nucleons of the target
nucleus. It has been demonstrated in \cite{KhT,Tuchin:2007pf} that
this is indeed the case for charm quark production in central and
forward \footnote{By forward rapidities we mean   the direction of the
  projectile fragmentation.} rapidities at RHIC and LHC.  

We assume in \eq{IE5} that the initial $c \bar{c}$ pair (with
transverse momentum $\un l_1$ in \fig{ccpp}) is colorless.  Indeed,
for large values of $k$ we can view the result of our calculation as a
product of two factors: the probability to find a gluon ($\un l_1$) in the
projectile hadron and its structure function in the target
nucleus. The gluon structure function can be modeled by the
interaction of a colorless probe such as dilaton or graviton
(\cite{dipole,AGL})  with the nucleus through the splitting into the
colorless $c \bar{c}$ pair. In Appendix we present a formal derivation
of all the main results of this section by direct summation of the
corresponding Feynman diagrams in the light-cone perturbation theory
along the lines of the dipole model \cite{dipole,KOTU}.  

Doing integrals over the  longitudinal positions $z_i$ of nucleons and
summing over $n$ in \eq{IE5} we obtain 
\be
&& \int^{2R_A}_{0}\rho\, \hat\sigma_{in}(x_2,r,r')\,d
z_0\,e^{-[\sigma(x_2,r^2) +\sigma(x_2,r'^2)]\,\rho\,2\,R_A }
\,\sum^{\infty}_{n=0} 
\int^{2 R_A}_{z_0}\,d\,z_{1}\, \dots \int^{2 R_A}_{z_{n-2}
}dz_{n-1}\int^{2 R_A}_{z_{n-1}} d z_n\,
\rho^n\,\hat\sigma^n_{in}(x_2,r,r')\nonumber\\ 
&& =\exp \left\{ -[\sigma(x_2,r^2) +\sigma(x_2,r'^2)]
\,\rho\,2\,R_A\right\} \,\cdot\,\Lb
\,e^{\hat\sigma_{in}(x_2,r,r')\,\rho\, 2R_A} \,-\,1\Rb
\nonumber \\ 
&&=\exp \left\{ - \sigma[x_2, (\un{r} - \un{r}' )^2]\,\rho\,2R_A \right\}
\,-\,\exp\left\{-[\sigma(x_2,r^2)
\,+\,\sigma(x_2,r'^2)]\,\rho\,2\,R_A \right\} \label{IE6}\,. 
\ee
Using \eq{IE6} we can reduce \eq{IE5} to the following expression
\be\label{IE7}
&&\frac{ d \sigma_{in}(pA)}{d Y\, d^2k\, d^2 b }\,
= \,x_1G(x_1,m^2_c)
\int d^2\,r \,d z\,e^{ -i\frac{1}{2}\un{k}\cdot \un{r}}\,
\int d^2\,r' \,d z'\,e^{ i\frac{1}{2}\un{k}\cdot
  \un{r}'}\,\Phi_G(l_1,r,r',z)\nonumber\\
&&\times \,\Lb  \exp \left\{ - \sigma[x_2, (\un{r} - \un{r}')^2]\,\rho\,2R_A \right\}  \,-\,\exp\left\{-[ \sigma(x_2,r^2)
+\sigma(x_2,r'^2)]\,\rho\,2\,R_A \right\} \Rb\,.  
\ee
This equation accounts only for the inelastic interaction and the
physical meaning of \eq{IE7} is the cross section of all possible
inelastic interaction in which the $c \bar{c}$ pair is produced.  We
need to add the cross section  for the elastic production of the
quark-antiquark pair, which reads 
\be
&&\frac{ d \sigma_{el}(pA)}{d Y\, d^2k\, d^2
  b}\,=\,x_1G(x_1,m^2_c)\,\int \,d^2\,r \,e^{ -i\frac{1}{2}\un{k}\cdot \un{r}}
\,\int \,d^2\,r' \,e^{ i\frac{1}{2}\un{k}\cdot
  \un{r}'}\, \Phi_G(l_1,r,r',z=1/2)\nonumber \\
&& \times \left\{ 1-\exp[- \sigma(x_2,r^2)\,\rho\,2R_A] \right\}
\,\cdot\, \left\{ 1- \exp[ -\sigma(x_2,r'^2)\,\rho\,2R_A]
\right\}\,. \label{IE71} 
\ee
The sum of \eq{IE7} and \eq{IE71} gives
\be
&&\frac{ d \sigma_{tot}(pA)}{d Y\, d^2k\, d^2
  b}\,=\,x_1G(x_1,m^2_c)\,\int \,d^2\,r \,e^{ -i\frac{1}{2}\un{k}\cdot
  \un{r}}
\,\int \,d^2\,r' \,e^{ i\frac{1}{2}\un{k}\cdot
  \un{r}'}\, \Phi_G(l_1,r,r',z=1/2)\nonumber \\
&& \times \left\{ 1- \exp[ - \sigma(x_2,r^2)\,\rho\,2\,R_A] -
\exp[ - \sigma(x_2,r'^2)\,\rho\,2\,R_A] +
\exp [ - \sigma(x_2, (\vec{r} - \vec{r}' )^2)\,\rho\,2R_A ]\right\}\,.
 \label{IE72}
\ee

Introducing the \emph{quark} saturation scale $\mathcal{Q}^2_s$ (see
 \eq{gluon.sat} and \eq{quark.sat}) we can write 
\beq\label{paramet} 
\sigma(x, r^2)\,\rho\,2R_A=\frac{1}{4}\,r^2\,\mathcal{Q}^2_{s,A}(x)\,.
\eeq
 The form of  $\mathcal{Q}^2_s$ is determined by the phenomenology of  low $x$ DIS \cite{MOD,MOD1,MOD2,MOD3,Iancu:2003ge} and forward hadron
production in $p(d)A$ collisions
\cite{KKT2,Dumitru:2005kb,Dumitru:2005gt,Goncalves:2006yt}.
Introducing a new dimensionless variable $\un\zeta = m_c \,\un r$ we can rewrite
\eq{IE72} as
 \be
 \frac{ d \sigma_{tot}(pA)}{d Y \, d^2k\, d^2b} &=& \frac{1}{(2\pi)^3}\frac{\as}{m_c^2\pi}\,
x_1G(x_1,m^2_c)\,\int d^2 \zeta \,d^2\zeta'\,e^{i \un{k}
  \cdot(\un{\zeta} - \un{\zeta}')/(2m_c)} \nonumber\\
  &&
 \times \left[\frac{1}{2}
\frac{\un{\zeta}\cdot\un{\zeta}'}{\zeta\,\zeta'}
K_1(\zeta\,)K_1(\zeta') +K_0(\zeta)\,K_0(\zeta')\right]\nonumber\\ 
&&
\times \left\{ 1- \exp\Lb - \zeta^2\,\Q^2_{s}/4m_c^2\Rb -
\exp\Lb - \zeta'^2\,\Q^2_{s}/4m_c^2\Rb +
\exp \left[ -  (\un{\zeta} - \un{\zeta}' )^2\,\Q^2_{s}/4m_c^2\right]\right\}\,. \label{IE8}
\ee

 It was pointed out in Ref.~\cite{KT}  that the dominant contribution
 to the integrals on the r.h.s.\  of  \eq{IE8} is originating from the
 integration region  $r'\ll r\ll 1/m_c$, i.e.\ 
$\zeta'\ll \zeta \le 1 $ (or, equivalently,  $\zeta\ll \zeta' \le 1$). In
 this kinematic region \eq{IE8} reduces to the following expression 
\begin{eqnarray} \label{IE9}
 \frac{ d \sigma_{tot}(pA)}{d Y\, d^2k\, d^2b}&=& 
 \frac{1}{4\pi}\frac{\as }{m_c^2\pi}\,
 \,x_1G(x_1,m^2_c)\nonumber\\
 &&\times \int_0^\infty  d\zeta^2
 \,K_0(\zeta)\,J_0(k\zeta/2m_c) 
\int^{\zeta^2}_0 d \zeta'^2 \,K_0(\zeta') \left\{ 1-\exp [ -
\zeta'^2\,\Q^2_s(x_2)/4 m^2_c ] \right\}
\end{eqnarray}
In the saturation region $\Q_s \,\gg\,m_c$ the dipole scattering amplitude reaches its unitarity limit $1-e^{-\zeta'^2\,\Q^2_s(x_2)/4 m^2_c}\approx 1$. Therefore, the rapidity distribution becomes 
\beq \label{IE10}
 \frac{ d \sigma_{tot}(pA)}{d Y\,d^2k\, d^2b}\,\propto \,
x_1G(x_1,m^2_c)\,\sim\, \exp\Lb - \lambda\,Y\Rb\,,
\eeq
 while for the same process in hadron-hadron collisions we have (see \eq{IE1})
\beq \label{IE11}
 \frac{ d \sigma_{tot}(pp)}{d Y\, d^2k\, d^2b}\,\propto \,
x_1G(x_1,m^2_c)\,x_2G(x_2,m^2_c)\,\sim\,\mathrm{constant}(Y)\,,
\eeq
where we assumed that $xG(x,m^2_c)\propto 1/x^{\lambda}$ at low $x$ (which is true if $Y$ is not too close to the proton fragmentation region). It is clear that there is a substantial difference between the rapidity distribution in these two cases.

 \subsection{Nucleus-nucleus collisions in the KLN approach}\label{sec:KLN}
 
 Nucleus-nucleus interaction can be characterized by the saturation
 scale which depends on the properties of both nuclei. In the KLN approach
 it is assumed that multi-particle production is entirely determined by the saturation scales 
 of the colliding nuclei $Q^2_{s,A_1}(x_1)$ and $Q^2_{s,A_2}(x_2)$. In the spirit of this approach we will generalize \eq{IE72} to the case of nucleus-nucleus scattering
using the Kovchegov's conjecture \cite{Kovchegov:2000hz}. In
\cite{Kovchegov:1998bi} Kovchegov and Mueller noted that in order that
their calculation of gluon production in $pA$ collisions be
self-consistent, an entire class of  the final state interactions must
cancel out in  the light-cone gauge. Although they did not find a
physical reason for that, Kovchegov suggested that the same conclusion
may hold in $AA$ collisions as well. Using this assumption he derived an
expression for gluon production in heavy ion collisions in the
light-cone gauge. 
In Appendix we derive \eq{main2} along the same lines of
reasoning. Here we would like to review the main steps. 

 \FIGURE[ht]{
\centerline{\epsfig{file=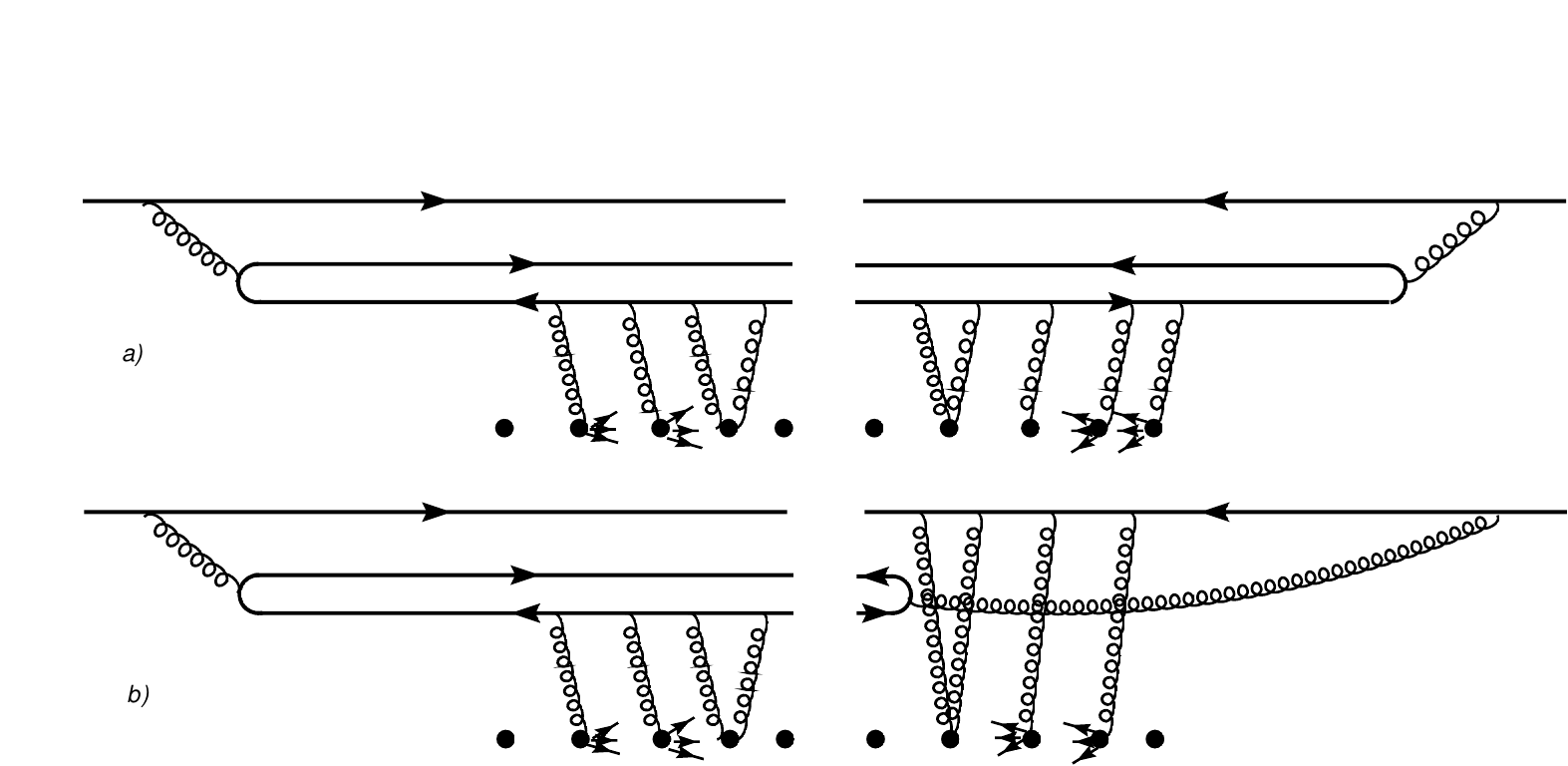,width=12cm}}
\caption{ The  main diagrams for the  process of inclusive $c \bar{c}$
  production  with fixed relative momentum in hadron-nucleus
  collisions (see Appendix).} 
\label{padia} }
The main contribution to the inclusive cross section of $c \bar{c}$
production stems from the diagrams shown in \fig{padia}. 
The sum of the diagrams of \fig{padia}-a are proportional to
\beq \label{PAD1}
x_1G(x_1,m^2_c)\,\left\{ e^{ -
    \frac{Q^2_{s,A}(x_2)}{8}\,(\un{r}\,-\,\un{r}' )^2}\,-\,
  1\,\right\}\,, 
\eeq
while the diagrams of  \fig{padia}-b are proportional to
\beq  \label{PAD2}
x_1G(x_1,m^2_c)\,\left\{ \Lb 1  \,-\, e^{ -
    \frac{Q^2_{s,A}(x_2)}{8}\,r^2}\Rb \,+\,\Lb 1  \,-\,e^{ -
    \frac{Q^2_{s,A}(x_2)}{8}\,r'^2} \Rb\right\} 
\eeq
 The sum of \eq{PAD1} and \eq{PAD2} gives \eq{IE72}.

\begin{figure}[ht]
\begin{center}
   \includegraphics[width=12cm]{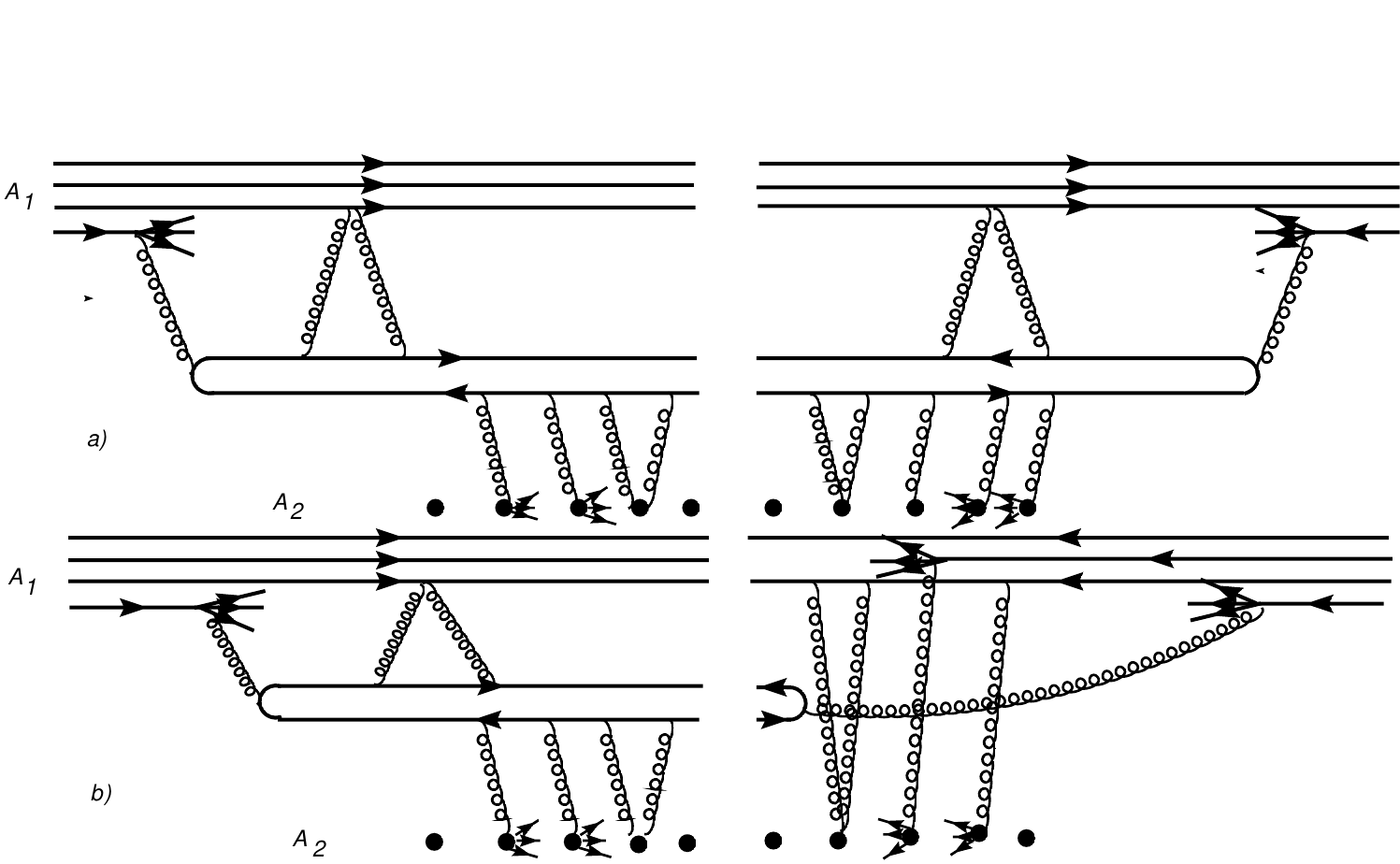} 
 \end{center}    
\caption{The main diagrams for the process of inclusive $c\bar c$
  production with fixed relative momentum in nucleus-nucleus
  collisions (see Appendix).} 
\label{aadia}
\end{figure}
For nucleus-nucleus collisions the main contribution stems from  the
set of the diagrams given in \fig{aadia}, which can be written as 
 \be
 && \frac{ d \sigma_{tot} (AA)}{d Y\, d^2k\, d^2b\,d^2b'}=
 \frac{1}{(2\pi)^3}\frac{\as }{m_c^2\pi}\,\frac{4N_c}{\as\pi^2}\nonumber\\
 &&
 \int \,d^2 \zeta \,d^2 \zeta'\,e^{i \vec{k} \cdot(\vec{\zeta} -
   \vec{\zeta}')/2m_c} \Lb\frac{1}{2}\,
 \frac{\un{\zeta}\cdot\un{\zeta}'}{\zeta\,\zeta'}
 K_1(\zeta\,)K_1(\zeta') \,+\,K_0(\zeta)\,K_0(\zeta')\Rb
 \label{IE13}\\ 
&&
\times \,\left(\,\frac{1}{\zeta^2}\, \left\{ 1- \exp[ -
  \zeta^2\,Q^2_{s,A_1}/8m^2_c] \right\}\,\left\{ 1- \exp[ -
  \zeta^2\,Q^2_{s,A_2}/8m^2_c] \right\}\,\right.\nonumber \\ 
&&
\left. +\,\frac{1}{\zeta'^2}\,\left\{ 1- \exp[ -
  \zeta'^2\,Q^2_{s,A_1}/8m^2_c] \right\}\,\left\{ 1- \exp[ -
  \zeta'^2\,Q^2_{s,A_2}/8m^2_c] \right\}\,\right.\nonumber \\ 
&&
\left. -\,\frac{1}{( \un{\zeta} - \un{\zeta}')^2}\,\left\{ 1-
  \exp[ - (\un{\zeta} -\un{\zeta}')^2\,Q^2_{s,A_1}/8m^2_c]
  \right\}\,\left\{ 1- \exp[ -(\un{\zeta} -\un{\zeta}')^2
  \,Q^2_{s,A_2}/8m^2_c]\right\} \,\right)\nonumber 
\ee
One can see that the first two terms in \eq{IE13} are the same as
\eq{PAD2} where factor $x_1G(x_1,m^2_c)$ is replaced by (see \eq{replace})
\beq \label{RPLCE1}
\frac{\as \pi^2}{4N_c} x_1G(x_1,m^2_c)\,\rightarrow\,\frac{d^2b}{r^2}\,\Lb
1 - e^{ - \frac{r^2\,Q^2_{s,A_1}}{8} } \Rb 
\eeq
or
$$
 \frac{\as   \pi^2}{4N_c}\,x_1G(x_1,m^2_c)\,\rightarrow\,\frac{d^2b}{r'^2}\,\Lb 1-
 e^{- \frac{r'^2\,Q^2_{s,A_1}}{8}} \Rb 
$$
  while the last term in  \eq{IE13} is equal to \eq{PAD1} with the replacement
\beq \label{RPLCE2}
\frac{\as \pi^2}{4N_c}
x_1G(x_1,m^2_c)\,\longrightarrow\,\frac{d^2b}{(\un{r} - \un{r}'
  )^2}\,\Lb 1-e^{-\frac{(\un{r} - \un{r}'
  )^2\,Q^2_{s,A_1}}{8} } \Rb 
\eeq
To understand this replacement we notice that the last two lines  in \eq{IE5}  can be written as 
\be \label{SUM2}
&&\int^{2R_A}_{0} \rho\,x_2\,G(x_2,m^2_c)\,d
z_0\,e^{-[\sigma(x_2,r^2) + \sigma(x_2,r'^2)]\,\rho\, (2 R_A - z_0) }
\, 
\sum^{\infty}_{n=0} 
\int^{2 R_A}_{z_0} \!d\,z_{1}\, \dots \int^{2
  R_A}_{z_{n-2}} \!dz_{n-1}\int^{2 R_A}_{z_{n-1}}  \!d z_n\,
\rho^n\,\hat\sigma^n_{in}(x_2,r,r')\nonumber  \\ 
&&= \int^{2R_A}_{0}\rho\,x_2\,G(x_2,m^2_c)\,d z_0\,\exp\bigg\{
 - \Big[\sigma(x_2,r^2) \,+\,\sigma(x_2,r'^2)  -\hat\sigma_{in}(x_2,r,r')\Big]\,\rho\,
(2R_A-z_0) \bigg\}\rightarrow  \\
&&    
 \frac{x_2\,G(x_2, 4/[\un{r} - \un{r}' ]^2)}{
  \sigma(x_2,r^2) \,+\,\sigma(x_2,r'^2)  -\hat\sigma_{in}(x_2,r,r')}\, 
\, \bigg( 1\,-\,\exp\bigg\{ - 
\Big[\sigma(x_2,r^2) \,+\,\sigma(x_2,r'^2)  -\hat\sigma_{in}
(x_2,r,r')\Big]\,\rho\,2\,R_{A} \bigg\} \bigg) \,\nonumber \\
&&=\frac{N_c}{\as \pi^2}\,\frac{1}{(\un{r} - \un{r}'
  )^2}\,\left\{\,1\,-\,\exp\left[ - \frac{(\un{r} - \un{r}' )^2 
\,Q^2_{s,A_1}(x_2)}{8}\right]
\right\} \,.
\ee
This corresponds to the diagram of \fig{aadia}-a. 
The sum in \eq{SUM2} reflects the fact that the dipole can scatter
elastically only after the first inelastic interaction. Contribution of the diagram 
\fig{aadia}-b is treated in the same way. The low density limits, i.e.\ hadron-hadron or hadron--nucleus collisions are reproduced when $Q^2_{s,A_1}\ll m_c^2$ and/or $Q^2_{s,A_2} \ll m_c^2$.

\FIGURE[ht]{
\centerline{\epsfig{file=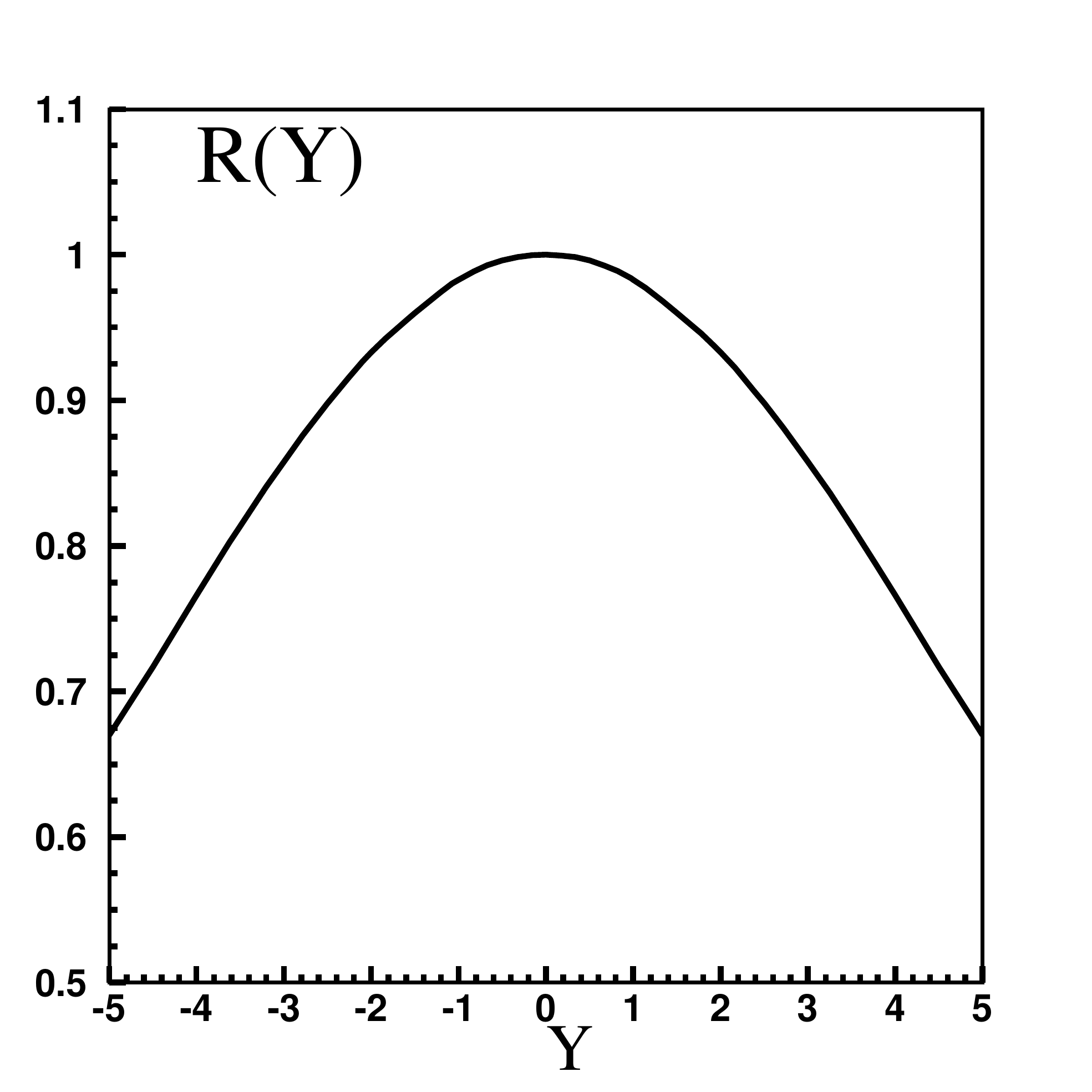,width=6cm}}
\caption{ The  ratio $R(y)$ defined in \eq{RATAA}, for the  process of
  inclusive $c \bar{c}$  production  with fixed relative momentum
  $\un{k}=0$)   in  nucleus-nucleus   collisions. The calculation is
  performed  for the Gold nuclei collision at RHIC energy $\sqrt{s} =
  200$~GeV using the KLN value for the saturation scale \cite{KLN}.} 
\label{aa-ym} }
In \fig{aa-ym} we plot the ratio
\beq \label{RATAA}
R( Y) \,= \,\dfrac{ \dfrac{d\sigma_{tot}(AA)}{dY\, d^2k\,
    d^2b}\Big|_{\un k=0}}{  \dfrac{d\sigma_{tot}(AA)}{dY\, d^2k\,
    d^2b}\Big|_{\un k=0,\,Y=0}}\,, 
\eeq
as function of  rapidity $Y$. This ratio has a much sharper maximum at $Y=0$ than the corresponding ratio in $pp$ collisions.

\section{$J/\psi$ production in hadron-nucleus collisions}

\subsection{New production mechanism off nuclear targets}

\FIGURE[ht]{
\centerline{\epsfig{file=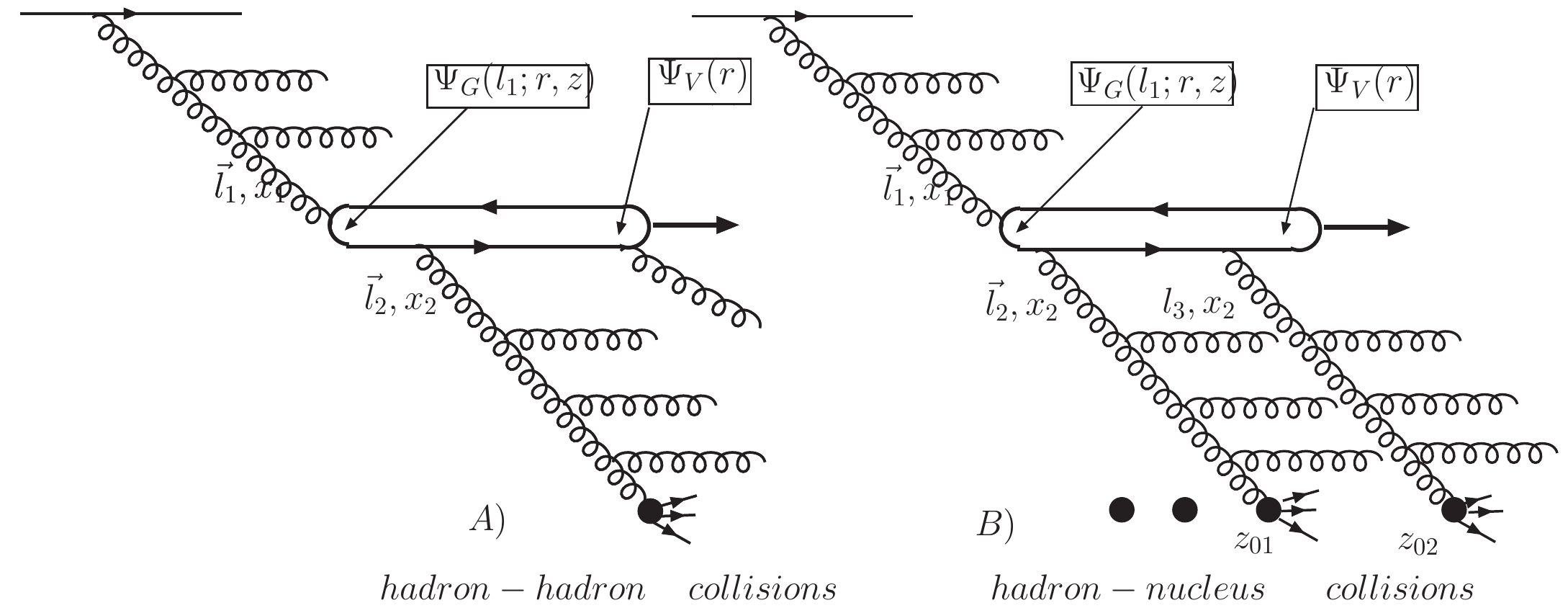,width=12cm}}
\caption{ The  process of inclusive  $J/\psi$  production in
  hadron-hadron (\protect\fig{psi1}-A) and  in hadron-nucleus
  collisions (\protect\fig{psi1}-B). } 
\label{psi1} }

The discussion and derivations of the previous section
now allow us to turn to the main subject of our paper. \fig{psi1} displays
the $ J/\psi$ meson production in $pp$ and $pA$ collisions at the leading
order in $\as^2 A^{1/3}$. The cross section for the latter is a direct
generalization of  \eq{IE1} and reads \footnote{All cross sections in
  this and the following sections are for the $ J/\psi$ production.} 
\be
  \label{PP1}
 \frac{d \sigma(pA)}{d Y\, d^2b }&\propto &
 \int\limits^{2 R_A}_0 \rho\,d z_{0}\,\int\limits^{z_{0}}_0
\rho\,d\,z_{1}\int\,\frac{d^2 l_1}{2 \pi}\,\phi_G(x_1,l_1)  \,
2\int d^2\,r \,d z\,\Psi_G(l_1,r,z)\otimes \Psi_V(r,z)\,\Lb 1 - e^{i \vec{l}_2
  \cdot \vec{r}} \Rb\,\Lb 1 - 
e^{i \vec{l}_3 \cdot \vec{r}} \Rb
\nonumber\\
&& \times \,2\,\int d^2\,r' \,d z'\,\Psi_G^*(l_1,r',z)\otimes \Psi_V^*(r',z')\Lb 1 -
e^{-i \vec{l}_2 \cdot \vec{r'}} \Rb\,\Lb 1 - 
e^{-i \vec{l}_3 \cdot \vec{r'}} \Rb
 \nonumber\\
&& \times \int\frac{d^2 l_2}{2 \pi\,l^2_2 }\,\phi_G(x_2,l_2)\,
 \int\frac{d^2 l_3}{2 \pi\,l^2_3 }\,\phi_G(x_2,l_3)
\ee
where $\Psi_G \otimes \Psi_V$ is projection of the $ J/\psi$ light-cone
``wave--function'' onto the virtual gluon one. Trace over all relevant quantum numbers is implied.  Assuming that
$\Psi_{V}(r,x)\,\propto \delta(z - 1/2)\,\delta(r)$ and $\un l^2_i\,
z(1 -z)\,\ll\,m^2_c$, $i=1,2,3$,  this projection takes the
following form \cite{KT,XS} 
\beq \label{WEV}
\Psi_{G}(m_c,r,z)\,\otimes
\Psi_V(r,z)\,=\,\sqrt{\frac{3\,\Gamma_{ J/\psi \to e^+
      e^-}\,M_{ J/\psi}}{48\,\pi\,\alpha_{em}}}\,\frac{m^3_c
  \,r^2}{4}\,K_2\Lb m_c\,r \Rb\,. 
\eeq

At short distances $r, r' \,<\,1/m_c$ the cross section in \eq{PP1} is proportional
to $r^2\,r'^2$. This fact reflects the higher twist nature of the
suggested mechanism. Since the position of the pair of nucleons is not
fixed the full contribution should be proportional to $A^{2/3}$  while
the mechanism 
of \fig{psi1}-A leads to an enhancement  by $A^{1/3}$.
The enhancement factor stems from integrations over $z_{0}$ and $z_{1}$ in \eq{PP1}
\beq \label{ENF}
\frac{d \sigma(pA)}{d Y\, d^2k}\,\propto\,  \int^{2 R_A}_0  \!\rho\,d z_{0}\,\int ^{z_0}_0  \rho\,d\,z_{1}\,=\,\frac{1}{2}\,\Lb \rho\,2 R_A \Rb^2\,.
\eeq
Parametrically, the mechanism in \fig{psi1}-B is different from that in \fig{psi1}-A by the factor $\as A^{1/3}$. Therefore, in the spirit of the quasi-classical approximation  in which we assume that $\as^2 A^{1/3}\sim 1$ we conclude that the mechanism in \fig{psi1}-B is enhanced by a big factor $1/\as$.

\subsection{Propagation of the colourless $c \bar{c}$ pair through a nuclear target}

In Ref.~\cite{KT}  detailed arguments were given which justify the 
application of the dipole model for calculation of $ J/\psi$ production
at  forward rapidities at RHIC. It has been argued that the
coherence length for the $c\bar c$ pair is sufficiently larger than
the longitudinal extent of the interaction region. This
means that the development of the light-cone ``wave function'' happens
a long time before the collision.  We relied on this physical picture
in our calculations in the previous section. Concerning $ J/\psi$, its
production is characterized by an additional formation time scale proportional to the
inverse binding energy $\sim \as^2\ m_c$; for a quantitative estimate, see \cite{Kharzeev:1999bh}. This time is certainly larger than the
charm quark pair production time (by a factor of about $1/\as^2$) implying that the
formation process takes place far away from the
nucleus. Therefore, in the following, we concentrate on the dynamics
of $c \bar{c}$ pair interactions with the nucleus. 

\FIGURE[h]{
\centerline{\epsfig{file=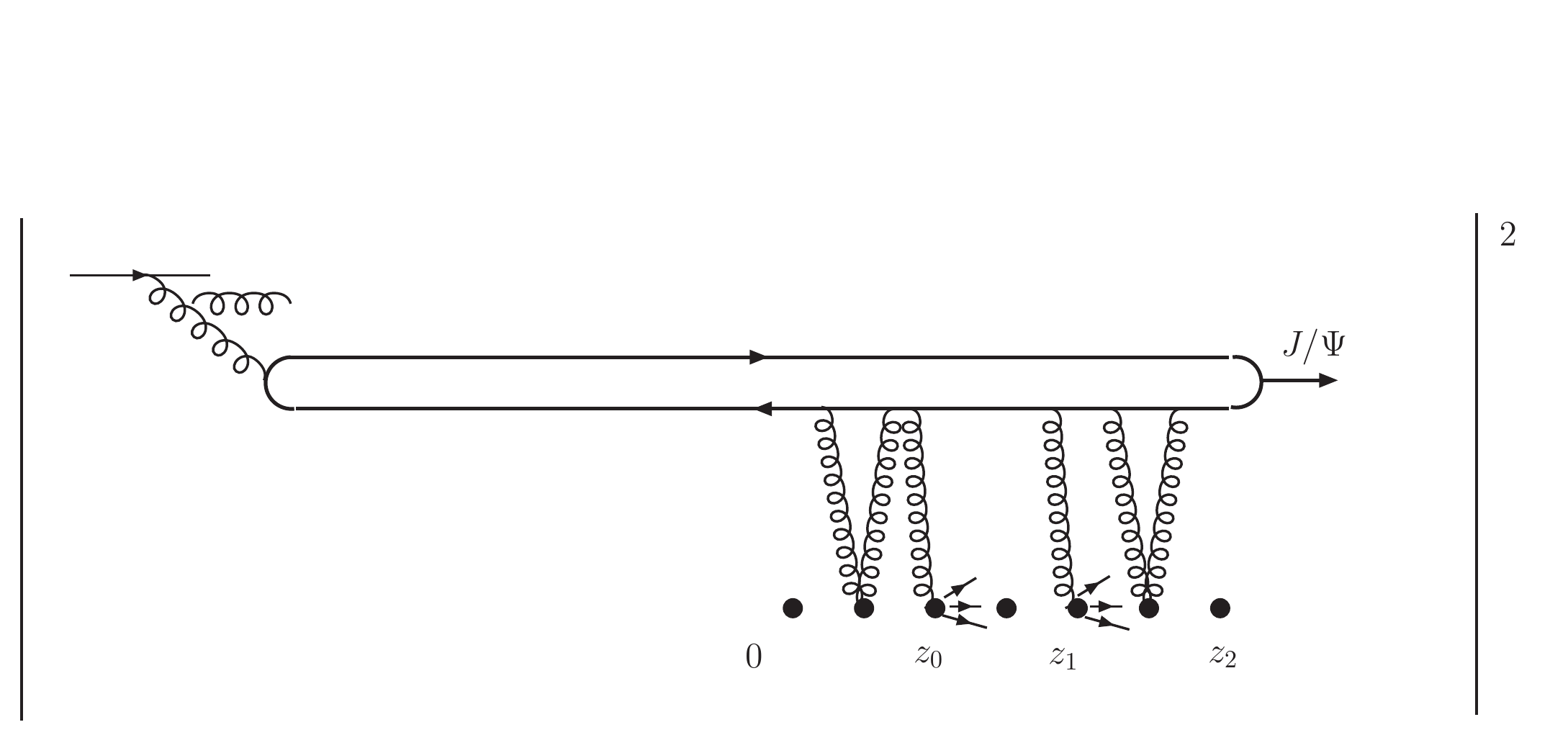,width=12cm}}
\caption{The  process of inclusive  $J/\psi$  production in
  hadron-nucleus collisions due to the interaction with an even number of
  nucleons.} 
\label{psi2} }
Since the soft gluon emission processes are suppressed in the
quasi-classical approximation, the $ J/\psi$ meson is predominantly
produced through the hadronization of the color \emph{singlet} $c\bar
c$ pair.  This rules out diagrams of the type \fig{padia}-b
corresponding to the elastic interaction. The diagrams of the type
\fig{padia}-a are shown in \fig{psi2}. Note that  since the $ J/\psi$
quantum numbers are $1^{--}$ while those of gluons are $1^-$ an odd
number of gluons must connect to the charm quark line. Consequently,
each inelastic interaction of  the $c \bar{c}$ pair must involve two
nucleons. To take this into account we write an analogue of  
\eq{IE5} in which the sum over all inelastic processes (i.e.\ sum over
$n$) involves only even number of interactions. We have 
\be
 \frac{d \sigma_{in}(pA)}{d Y\, d^2b}& =&  \, C_F\, x_1
G(x_1,m^2_c) \,\label{PP3}\\ 
&&\times\int^{2 R_A}_0 \rho\,\hat\sigma_{in}(x_2,r,r')\,d\,z_{0}
\,\int \,d^2\,r \,\Psi_G(l_1,r,z=1/2)\,\Psi_V(r)\,\otimes\,\int\,d^2
r' \Psi^*_G(l_1,r',z=1/2)\,\Psi^*_V(r')\nonumber \\ 
&& \times  \Lb\,e^{-(\sigma(x_2,r^2)
  \,+\,\sigma(x_2,r'^2))\,\rho\,2\,R_A } \,\sum^{\infty}_{n=0} 
\int^{2 R_A}_{z_0}\,d\,z_{1}\, \int^{2 R_A}_{z_1} d z_2 \dots\int^{2
  R_A}_{z_{2n}} d z_{2n +1} \,\rho^{2n+1}\, 
\hat\sigma^{2n+1}_{in}(x_2,r,r')\Rb \nonumber
\ee
The color factor in front comes from the calculation of \fig{psi1}-B,
namely, it is equal to 
\be \label{CC}
 \Tr (t^a t^b t^c)\,\Tr(t^a t^b t^{c'})\,\delta^{c
   c'}&=&\frac{1}{16}\,\delta^{c c'}\,\Lb
 f_{abc}\,f_{abc'}\,+\, d_{abc}\,d_{abc'} \Rb\\ 
&&\hspace{-1cm}=\frac{(N^2_c -1)}{16 N_c}\,\Lb N_c + \frac{N^2_c -4}{N_c}
\Rb  \,= \frac{N_c^2-1}{2N_c}\,\frac{N_c^2-2}{4}\,=\left(
  \frac{N_c^2-1}{2N_c}\right)^2\,\frac{N_c(N_c^2-2)}{2(N_c^2-1)}\approx C_F^3\,. 
\nonumber 
\ee
Since $\hat{\sigma}_{in}$ is proportional to $C_F=(N^2_c-1)/2 N_c$, we
extract this factor from the color coefficient of \eq{CC}. The last of
equations in \eq{CC} is written in the large $N_c$ approximation.  

 We argued in Sec.~\ref{sec:KLN} (see \fig{padia}) and in the Appendix that 
 $c \bar{c}$ pair in the color octet  state passes through the target
 with the same  elastic  (see \fig{padia}-b and  \eq{PAD2}) and
 inelastic  (see \fig{padia}-a and  \eq{PAD1}) cross sections as the
 $c\bar c$ pair in the color singlet state. This is the reason we
 do not  need to change \eq{PP3}  to include the  color octet state
 interaction with the target.

After integration over $z_i$'s  and summation over $n$ using the identity 
$\sum_{n=0}^\infty   a^{2n+2}/(2n+2)!=\cosh a-1$ we obtain the
following formula 
 \be
 \frac{d \sigma_{in}(pA)}{d Y\, d^2b} &= & C_F\, x_1
 G(x_1,m^2_c) \, \label{PP5} 
 \int d^2\,r
\Psi_G(l_1,r,z=1/2)\,\Psi_V(r)\int\,d^2 r'
\Psi^*_G(l_1,r',z=1/2)\,\Psi^*_V(r') \nonumber \\
&\times&  \frac{1}{2}\bigg\{ \exp\Big[ - \sigma\Lb x_2,(\un{r} -
  \un{r}')^2\Rb\,\rho\,2R_A\Big] \bigg. 
 \bigg.+ \exp\Big[ -
( \sigma(x_2,r) + \sigma(x_2,r') +\hat
\sigma_{in}(x_2,r,r'))\,\rho\,2R_A\Big] \nonumber \\ && 
\hspace{5cm}
-2\,\exp\Big[ -
(\sigma(x_2,r) + \sigma(x_2,r'))\,\rho\,2 R_A\Big] \bigg\}~.
 \ee
The color factor in \eq{PP3} as well as in \eq{PP5} corresponds to the
diagram of \fig{psi1}-B.

 \subsection{ J/${\bm \Psi}$ production in hadron-nucleus collisions
   in the saturation regime}\label{sec:3.3} 
 
In the quasi-classical approximation the  gluon saturation scale is given by
\cite{dipole,AGL}, see \eq{gluon.sat}  
 \beq \label{QS}
 Q^2_{s,A}( x)\,=\,4\,\pi^2\as^2\,\rho\,T(\underline{b})\,,
 \eeq
 where $\rho$ is the nucleon density in a nucleus, $N_c$ is the number
 of colours, $\un b$ is the impact parameter and $T(b)$ is the optical
 width of the nucleus.  Eq.~\eq{QS} determines the scale of the
 typical transverse momenta  for the inclusive gluon production
 \cite{KOTU}. Its value was extracted from the fit   
 to the hadron multiplicities in nuclear collisions at RHIC \cite{KN1,KLN}. 
 However, for the penetration of the quark-antiquark pair  the typical
 saturation scale is about twice as small and we refer to it as the
 quark saturation scale $\Q^2_{s,A}$, see \eq{quark.sat}. This scale
 was extracted from fits of the  $F_2$ structure function in DIS
 \cite{AGL,MOD,MOD1,MOD2,MOD3} as we have already mentioned.  Both phenomenological approaches
 agree with each other, so the use of either quark or a properly rescaled gluon saturation scale is merely a matter of 
 convention.  In this paper we will use the gluon saturation scale
 \eq{QS}. Using this definition for the saturation momentum we have 
 $\sigma(x_2,r)= r^2 Q^2_s(x_2)/8$ (cf.\ \eq{paramet}).  Substituting this
 expression into \eq{PP5} we can re-write it in a more convenient form 
\be
 \frac{d \sigma_{in}(pA)}{d Y\, d^2b} \,&=& \,C_F\, x_1 G(x_1,m^2_c) \, \label{PP51} \\
 &&
\times \int d^2\,r\,  \Psi_G(l_1,r,z=1/2)\otimes\Psi_V(r)\int\,d^2
r'\, \Psi_G^*(l_1,r',z=1/2)\otimes\Psi^*_V(r') \nonumber\\ 
&\times&  \frac{1}{2}\left\{ \exp\Lb - \frac{(\vec{r} -
    \vec{r}')^2\,Q^2_{s,A}}{8}\Rb \,+\, \exp\Lb - 
\frac{(\vec{r} + \vec{r}')^2\,Q^2_{s,A}}{8}\Rb \,-\,2\,\exp\Lb -
\frac{(r^2 + r'^2)\,Q^2_{s,A}}{8}\,\Rb \right\} \nonumber\,. 
 \ee
Integrating over the angle between $\un{r}$ and $\un{r}'$ we derive
\be
 \frac{d \sigma_{in}(pA)}{d Y\, d^2b} \,&=& \,\frac{N_c (N^2_c -
   2)}{2\,(N^2_c - 1)}\, x_1 G(x_1,m^2_c)  
\label{PP6} \\
 &&
 \times\,\int d^2\,r \,
 \Psi_G(l_1,r,z=1/2)\,\Psi_V(r)\,\otimes\int\,d^2 r'\,
 \Psi_G^*(l_1,r',z=1/2)\,\Psi^*_V(r') \nonumber\\ 
&\times & \,  \exp\Lb - \frac{(r^2+ r'^2)\,Q^2_{s,A}}{8}\Rb
\,\left\{\, I_0\Lb \frac{Q^2_{s,A}}{4}\,r\,r'
  \Rb\,-\,1\,\right\}\,. \nonumber 
 \ee
  
 Deeply in the saturation region where $Q_{s,A}\gg\,m_c$ the typical
 dipole sizes are much smaller than $1/m_c$. Thus, we can expand the
 wave function \eq{WEV} $\Psi_G\otimes \Psi_V \approx \mathrm{const}$.
 The main contribution comes from $(\vec{r} - \vec{r}')^2  \leq
 1/Q^2_{s,A}$ while $ r \approx 1/m_c$. It gives 
\beq \label{PP61}
 \frac{d \sigma_{in} \Lb pA \Rb}{d Y d^2 b} \,\propto x_1G(x_1, m^2_c
 )/Q^2_{s,A}(x_2)\,\propto\,\exp\Lb - 2\,\lambda Y \Rb\,. 
 \eeq
In deriving \eq{PP61} we used the same assumptions as 
 in the case  of $c \bar{c}$-pair production with fixed
 relative momentum (see \eq{IE10}).  One can see that \eq{PP61} leads
 to a rapidity distribution that is more narrow than the distribution in
hadron--hadron collisions  given by \eq{IE11}.

 \section{Inclusive  $J/\psi$ production in nucleus--nucleus collisions}

 Using the same arguments as in Sec.~\ref{sec:KLN} which led us to \eq{IE13} 
we can  generalize \eq{PP6} to obtain our main result -- the formula for  $J/\psi$
 production in nucleus--nucleus collisions.  
 It reads 
 \be
 &&\frac{1}{S_A}\frac{d \sigma(AA)}{d Y\, d^2b} \,=  \,
 \frac{C_F^2}{4\pi^2\as} \int d^2 r
 \,\Psi_G(l_1,r,z=1/2)\otimes\Psi_V(r)\int\,d^2 r'
 \,\Psi^*_G(l_1,r',z=1/2)\otimes\Psi^*_V(r') \label{IP1} \\ 
 &&\times\, \frac{1}{2 \un{r} \cdot \un{r}'} \left\{  \exp\Lb -
   \frac{1}{8}(\un{r}  - \un{r}')^2 \, (Q^2_{s,A_1}+Q^2_{s,A_2}) \Rb
   \,-\,\exp\Lb - \frac{1}{8}(\un{r}   
 + \un{r}')^2 \, (Q^2_{s,A_1}+Q^2_{s,A_2}) \Rb\,  \right. \nonumber\\ 
&& \left . -\,\exp\Lb -\frac{1}{8}
 (\un{r}  - \un{r}')^2 \, Q^2_{s,A_1} -\frac{1}{8}(r^2 +
 r'^2)\,Q^2_{s,A_2} \Rb\,+\, \exp\Lb -\frac{1}{8} (\un{r}  +\un{r}')^2 \,
  Q^2_{s,A_1} -\frac{1}{8}(r^2 + r'^2)\,Q^2_{s,A_2} \Rb\, \right. \nonumber \\ 
&& \,\left. -\,\exp\Lb
  -\frac{1}{8} (\un{r}  - \un{r}')^2 \, Q^2_{s,A_2} -\frac{1}{8}(r^2 +
  r'^2)\,Q^2_{s,A_1} \Rb\,
+\,\exp\Lb - \frac{1}{8}(\un{r}  + \un{r}')^2 \,
  Q^2_{s,A_2} -\frac{1}{8}(r^2 + r'^2)\,Q^2_{s,A_1} \Rb 
\right\}\,, \nonumber
\ee
where $S_A$ is the transverse overlap area. 
One can check that  this formula  describes the hadron-nucleus $ J/\psi$  assuming that $Q^2_{s,A_1}$ is small.

\FIGURE[h]{
\centerline{\epsfig{file=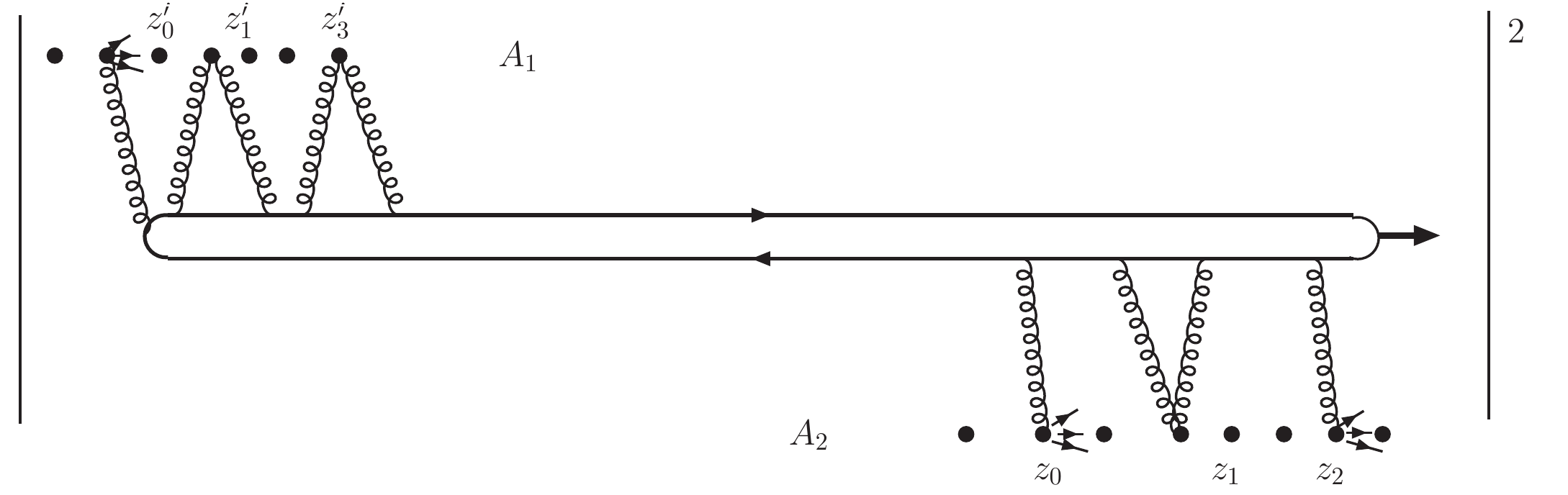,width=12cm}}
\caption{ The  process of inclusive  $J/\psi$  production in nucleus-nucleus collisions due to inelastic interaction  with both nuclei.
 }
\label{psiaa} }
In deriving \eq{IP1}  we  summed up the inelastic cross sections for
both nuclei. Let us denote the nucleon coordinates in the nucleus $A_1$ by
 $z'_i$ and in the nucleus $A_2$ by $z_i$. The number of possible
inelastic interactions (see \fig{psiaa}) is odd for both nuclei. For a
fixed number of total inelastic interactions $2n-1$ the number of
interactions in each nucleus can be $1\le k\le 2n-2$. Therefore, for
the term inside the curly brackets  in \eq{IP1} we have  
\be
\{\dots \} &=&\, 
\int^{2R_{A_2}}_{0} \,
\int^{2R_{A_1}}_{0}\Lb\frac{1}{8}\,Q^2_{s,A_2}\Rb\,\Lb\frac{1}{8}\,Q^2_{s,A_1}\Rb
\,(2\un{r}\cdot\un{r}')^2 \,d z_{0}\,d z'_0\,\exp\left\{- \frac{1}{8}\,( r^2
  + r'^2)\,(Q^2_{s,A_1}+Q^2_{s,A_2})\right\}  \,\nonumber\\ 
&&\times\,\sum^{2n-2}_{k=1}
\int^{2 R_{A_2}}_{z_0}\,d\,z_{1}  \int^{2 R_{A_2}}_{z_1} d z_2\dots
\int^{2 R_{A_2}}_{z_{k-2}} d z_{k-1}\,
\rho^{k}\,\Lb\frac{1}{8}\,Q^2_{s,A_2}\,2\,\un{r}\cdot\un{r}'\Rb^{k-1}\nonumber
\\ 
  & \times & \sum^{\infty}_{n =2} 
\int^{2R_{A_1}}_{z'_0}\,d\,z'_{1}  \int^{2R_{A_1}}_{z'_1} d z'_2\dots
\int^{2R_{A_1}}_{z'_{2n-k -2}} d z'_{2n - k -1}\, \rho^{2n -k-1
}\,\Lb\frac{1}{8}\,Q^2_{s,A_1}\,2\,\un{r}\cdot \un{r}'\Rb^{2n - k -2}
\label{IP11} 
\ee
Using the following mathematical identity
\beq \label{SF}
\sum^{j-1}_{k=1} \frac{1}{k! (j -k)!}\,a^k\,b^{j - k}\,=\,
\frac{1}{j!}\,( a+b)^j\,-\,\frac{1}{j!} a^j\,-\,\frac{1}{j!} b^j  
\eeq
with $j=2n-1$ we obtain 
\be\label{88}
&&\sum_{n=2}^\infty \left\{ \frac{1}{(2n-1)!}
\Lb\frac{1}{8}\,(Q^2_{s,A_1}+Q^2_{s,A_2})\,2\,\un{r}\cdot\un{r}'\Rb^{2n-1}-
\frac{1}{(2n-1)!}
\Lb\frac{1}{8}\,Q^2_{s,A_1}\,2\,\un{r}\cdot\un{r}'\Rb^{2n-1}\right.\nonumber\\
&&
\left. -
\frac{1}{(2n-1)!}
\Lb\frac{1}{8}\,Q^2_{s,A_2}\,2\,\un{r}\cdot\un{r}'\Rb^{2n-1}\right\}\nonumber\\
&&= \sinh\Lb\frac{1}{8}\,(Q^2_{s,A_1}+Q^2_{s,A_2})\,2\,\un{r}\cdot\un{r}'\Rb -
\sinh\Lb\frac{1}{8}\,Q^2_{s,A_1}\,2\,\un{r}\cdot\un{r}'\Rb
-\sinh\Lb\frac{1}{8}\,Q^2_{s,A_2}\,2\,\un{r}\cdot\un{r}'\Rb\,.
\ee
yielding \eq{IP1}. If $Q_{s,A_2}^2\ll Q_{s,A_1}^2$ we can expand \eq{88} using $\sinh(a+b)-\sinh a-\sinh b \approx b\,(\cosh a-1)+ \mathcal{O}(b^2)$; then  \eq{IP1}  reduces to \eq{PP51}.

For a qualitative discussion it is instructive to rewrite \eq{IP1} in
the region $r' \ll r\approx 1/Q_{s,A} \ll 1/m_c$. Expanding expression
in the curly brackets we derive  
\beq
\{\dots\}=\frac{1}{64}
\,Q_{s,A_{1}}^2\, Q_{s,A_{2}}^2\, (Q^2_{s,A_1}+Q^2_{s,A_2})\, (\un r\cdot \un
r')^3\,\Lb1\,+\,\mathcal{O}(Q_{s,A}^2r'^2\,, Q_{s,A}^4r'^2r^2)\Rb 
\, \exp\Lb-\frac{1}{8}(Q^2_{s,A_1}+Q^2_{s,A_2})\, r^2\Rb\,.
\eeq
In this approximation Eq.~\eq{IP1} becomes (after integration over the
angle between $\un r$ and $\un r'$) 
 \be
  \frac{1}{S_A}\frac{d \sigma(AA)}{d Y\, d^2b}  &\propto& \, \int
 d^2 r\, \Psi_G(l_1,r,z=1/2)\otimes\Psi_V(r)\int\,d^2 r'\,
 \Psi_G^*(l_1,r',z'=1/2)\otimes\Psi^*_V(r') 
 \nonumber \\
&& \times 
  Q^2_{s,A_1}\,Q^2_{s,A_2}\,(Q^2_{s,A_1}+Q^2_{s,A_2})\,r^2\,
r'^2\,  \exp\left\{- r^2\,(Q^2_{s,A_1}+Q^2_{s,A_2})/8\right\}\label{IP32}\\ 
&\propto& \frac{Q^2_{s,A_1}\,Q^2_{s,A_2}}{(Q^2_{s,A_1}+Q^2_{s,A_2})^3}\,. \label{IP3}
\ee
In \eq{IP3}  we replaced the wave functions of \eq{WEV} by a constant
as already discussed in Sec.~\ref{sec:3.3} and took the
integral over the angle between $\vec{r}$ and $\vec{r}'$. 

 From \eq{IP3} one can see that the spectrum of $J/\psi$'s in
 ion-ion collisions is more narrow than the one in hadron-hadron.  Explicitly  
 \beq  \label{IP31}
 \frac{d \sigma (AA)}{d Y}\,\propto\, \frac{d \sigma (pp)}{d
   Y}\frac{1}{(Q^2_{s,A_1}+Q^2_{s,A_2})^3} 
 \,\propto\, \frac{d \sigma (pp)}{d Y}\,e^{ -3 \lambda |Y|}\,,
 \eeq
where we use that $Q^2_s(x) \propto (1/x)^\lambda$ and $Y$ is the
rapidity of $J/\psi$ in the center-of-mass frame. 
 Therefore, our prediction is that the rapidity distribution of
 $J/\psi$ is much more narrow in nucleus-nucleus collisions than in
 the proton--proton ones. 

 To evaluate how close we  are to the saturation region at RHIC
 energies in this process we  first rewrite the general formula for
 the kinematic region $r \,\gg\,r'$. It takes the following form: 
 \beq
\frac{1}{S_A}\frac{d \sigma(AA)}{d Y\, d^2b} \propto
 Q^2_{s,A_1}(x_1)\,Q^2_{s,A_2}(x_2)\,[Q^2_{s,A_1}(x_1)+
 Q^2_{s,A_2}(x_2)]\,  \int_0^\infty d \zeta\, \zeta^9
 \,K_2(\zeta)\,e^{- \frac{\zeta^2}{8\,m^2_c}\,\left[
   Q^2_{s,A_1}(x_1)\,+\,Q^2_{s,A_2}(x_2)\right] } \label{IP4} 
 \eeq

\begin{figure}
\begin{center}
\includegraphics[width=6cm]{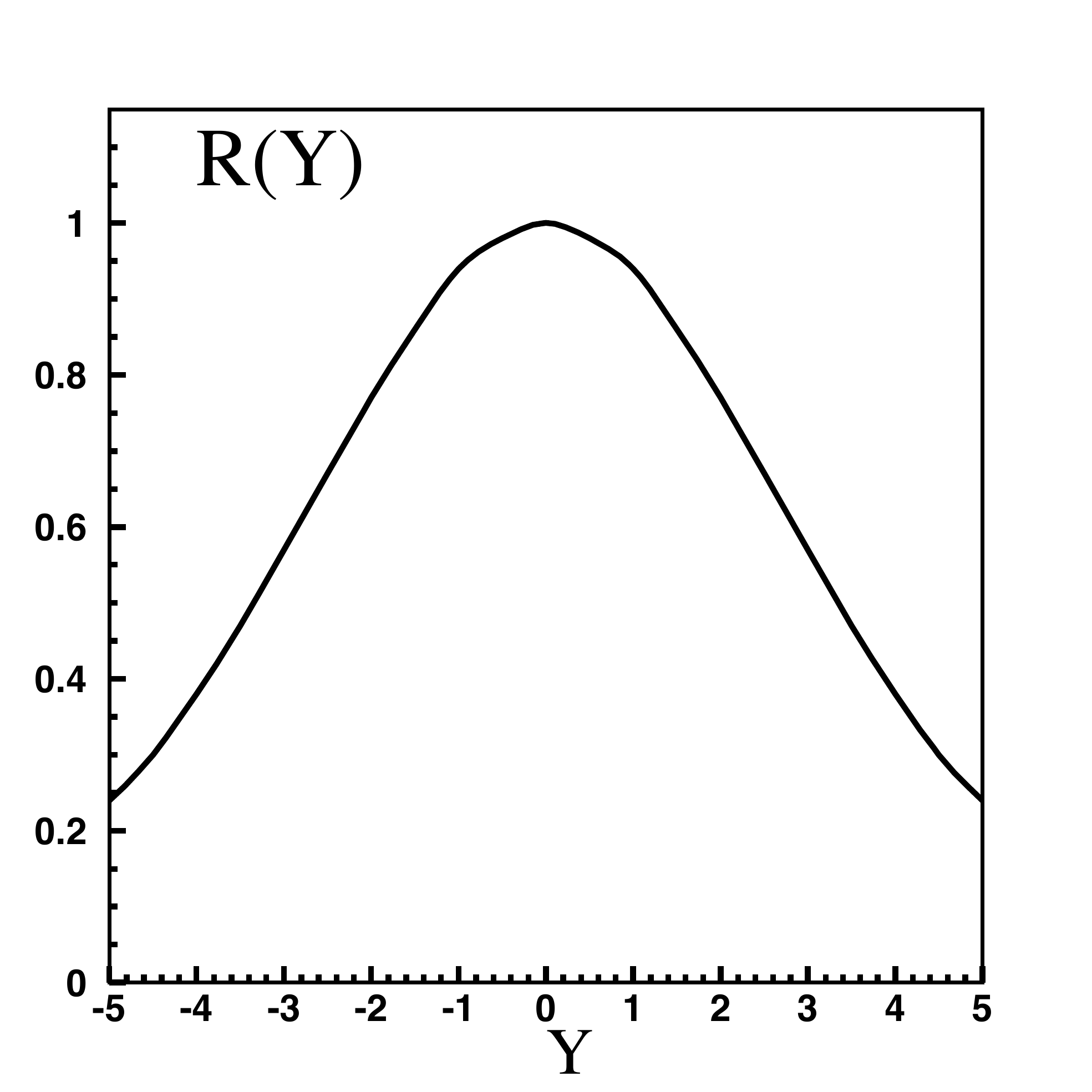}
\caption{Rapidity  dependence of  the ratio 
$R(Y)=\frac{\frac{d \sigma}{d Y}(Y)}{\frac{d
    \sigma}{d Y}(Y=0)}$  for the gold-gold collision
  at RHIC. For the saturation momenta the KLN expression was used. }
\label{q}
\end{center}
\end{figure}


In \fig{q}  we plot  the result of our calculation  for the ratio 
$R(Y) \equiv \frac{d \sigma}{d Y}(Y) / \frac{d
    \sigma}{d Y}(Y=0)$
using
\eq{IP4}. 
  For $Au-Au$ collision at RHIC with $\sqrt{s}=200$ GeV taking the KLN value for the
  saturation momentum $Q^2_s(y=0) = 2.2$~GeV$^2$ 
  for central collisions we find  that the rapidity distribution turns out
  to be very narrow although not quite to an extent suggested by the approximate expression 
  \eq{IP31} (see \fig{q}).  The rapidity distribution in \fig{q}
  is driven by the ratio $ Q^2_{s,A_1}\,Q^2_{s,A_2}/(Q^2_{s,A_1}+Q^2_{s,A_2})^3$.  The cross section decreases with the increase
  of the value of the saturation momentum $Q^2_{s,A_1}+Q^2_{s,A_2}$. However, this decrease is
  much milder than in \eq{IP31}.

\section{Numerical calculations}\label{sec:numerics}

In this section we perform numerical calculations of inclusive
$ J/\psi$ production using  \eq{IP1}. First of all, we reinstall  
the impact parameter dependence of the saturation scales and consider
a realistic distribution density for nuclei. Recall that
$Q_s^2\propto \rho T(b)$. Denote the impact parameter between centers
of two nuclei as $\un{b}$.  The position of a nucleon inside  nucleus
$A_1$ with respect to its center denote by $\un{s}$. Then the position
of a nucleon in the nucleus $A_2$ is given by $\un{b} - \un{s}$. We have 
\beq
Q_{s,A_{1}}^2\, \to\, Q_{s,A_{1}}^2(\un s)\,,\quad
Q_{s,A_{2}}^2\, \to\, Q_{s,A_{2}}^2(\un b-\un s)\,.
\eeq 
In our Glauber-type approximation (see e.g. \cite{Kharzeev:1996yx}) we neglect the impact parameter dependence in
nucleon-nucleon interactions considering their range much smaller than the size of
nuclei. The observable that we are going to
calculate is the number of $J/\psi$'s inclusively produced in nucleus--nucleus 
collisions at a  given rapidity $Y$ and a centrality characterized by the impact parameter 
$b$. The corresponding expression reads 
\be
\frac{d N^{AA}(Y,b)}{d Y} &\propto& \int d^2s\,
 Q^2_{s,A_1}(x_1,\un s)\,Q^2_{s,A_2}(x_2, \un b-\un
 s)\,[Q^2_{s,A_1}(x_1, \un s)+ Q^2_{s,A_2}(x_2,\un b-\un s)]\,
 \nonumber\\ 
 &&\times \int_0^\infty d \zeta\, \zeta^9 \,K_2(\zeta)\,\exp \left\{-
 \frac{\zeta^2}{8\,m^2_c}\,[ Q^2_{s,A_1}(x_1,\un
 s)\,+\,Q^2_{s,A_2}(x_2,\un b-\un s)] \right\}\,.\label{IP4a} 
 \ee
 where $\frac{d N^{AA}(Y,b)}{d Y}= \frac{d \sigma^{AA}(Y,b)}{d Y\, \sigma^{AA}_\mathrm{tot}}$ and 
\beq \label{XX}
  x_1\,=\,\frac{m_{J/\psi}}{\sqrt{s}}\,e^{-Y}\, ,\quad 
   x_2\,=\,\frac{m_{ J/\psi}}{\sqrt{s}}\,e^{Y}
\eeq

We can also write down \eq{IP4a} in the following way: 
 \be
 \frac{d N^{A A}(Y,b)}{ d Y}&=& C \frac{d N^{pp}(Y)}{d Y}\,\int d^2
 s\, \,T_{A_1}(\un s)\,T_{A_2} \Lb\un{b} - \un{s}\Rb\, 
[  Q^2_{s,A_1}\Lb x_1,\un{s}\Rb \,+\,Q^2_{s,A_2}\Lb x_2,\un{b} -
\un{s}\Rb] \,\frac{1}{m^2_c} 
 \nonumber\\
 & &
 \times  \int^{\infty}_0  d \zeta\,\zeta^9 \,K_2(\zeta)\,
 \,\exp\left\{ - \frac{\zeta^2}{8 m^2_c}\,[
 Q^2_{s,A_1}(x_1,\un{s})\,+\,Q^2_{s,A_2}(x_2, \un{b} - \un{s})]
 \right\}\,.\label{IP5}
  \ee
The overall normalization constant $C$  includes the color  and the
geometric factors $C^2_F/(4 \pi^2 \as S_p)$ where $S_p$ 
is interaction area in proton--proton collisions; it also  includes a rather poorly known amplitude of charm quark--antiquark transition into $ J/\psi$ and a
 gluon in the case of $pp$ collisions (see \fig{psi1}-A).  

\begin{figure}
\begin{center}
\includegraphics[width=9cm]{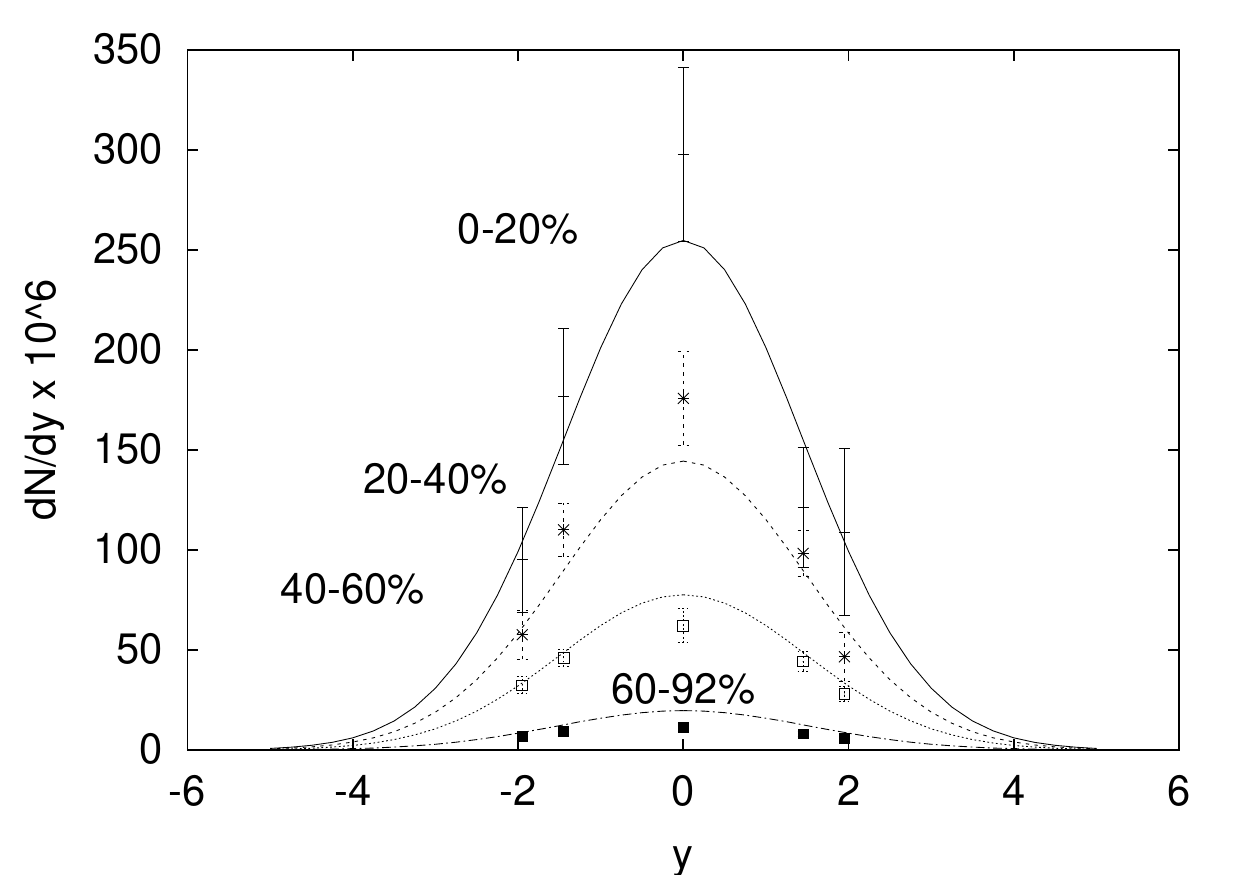}
\caption{$J/\psi$ rapidity distribution in Au-Au collisions for
  different centrality cuts. Experimental data from \cite{Adare:2006ns}.
   \label{fig:BdNdy}
}
\end{center}
\end{figure}

To calculate the multiplicity of $J/\psi's$ in $AA$ collisions using \eq{IP5} we need to know (i) rapidity distribution of $J/\psi$ multiplicity  ${d N^{pp}}/{d Y}$ in $pp$ collisions and (ii) the overall normalization constant $C$. 
We fitted the rapidity distribution of $J/\psi$'s in $pp$ collisions to
the experimental data of Ref.~\cite{Adare:2006kf} with a single
gaussian. The global normalization factor $C$ is
 found from the overall fit.

Now we can compute the nuclear modification of the $J/\psi$ rapidity distribution in $AA$ collisions at all centralities using \eq{IP5}. 
In figure \ref{fig:BdNdy}  we compare our results 
with the  experimental data of PHENIX Collaboration \cite{Adare:2006ns} for
Au-Au collisions at $\sqrt{s}=200$ GeV.  
The agreement between our calculation and the experimental data is reasonable.

\begin{figure}
\begin{center}
\includegraphics[width=12cm]{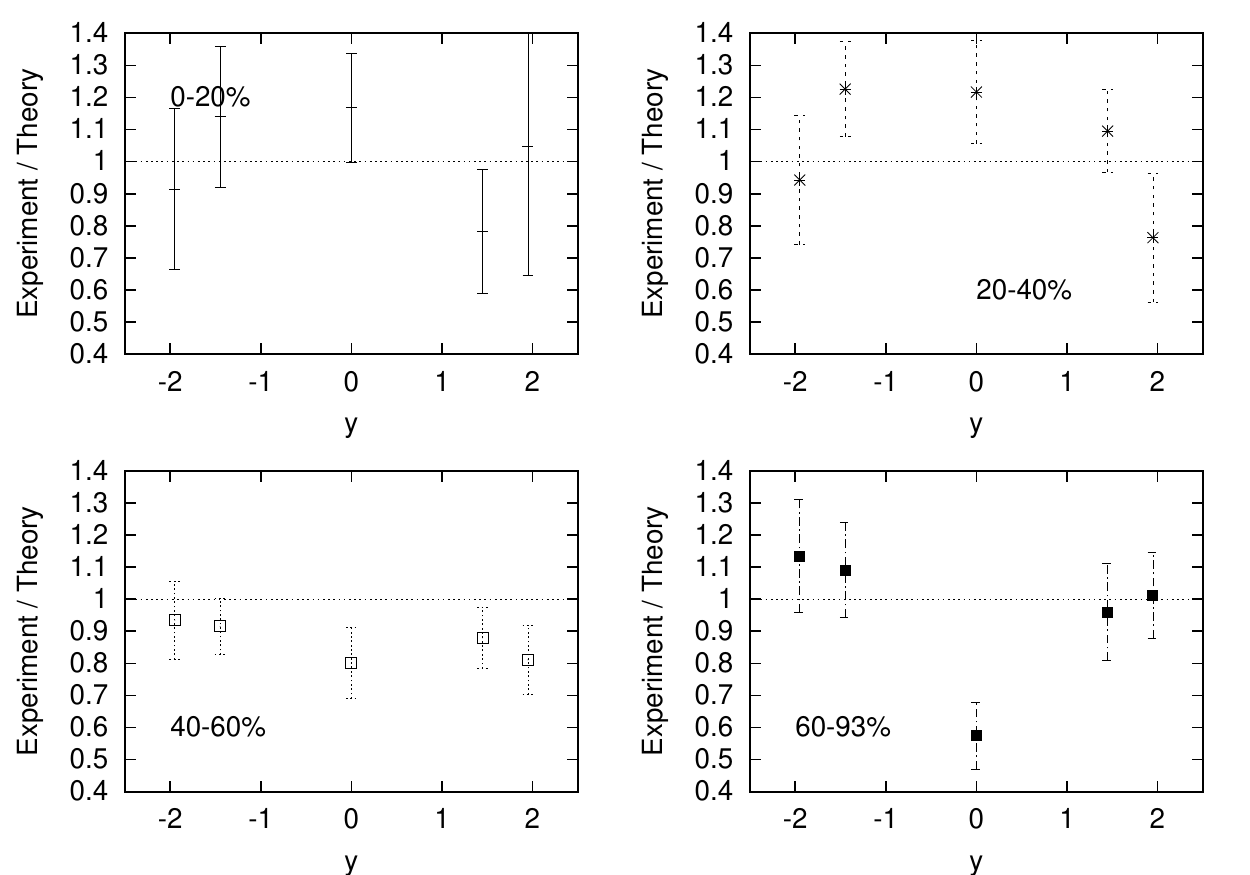}
\caption{Ratios of $J/\psi$ rapidity distribution in $Au-Au$ collisions:
Experimental data \cite{Adare:2006ns} divided by theoretical our results.  }     \label{fig:BdNdy_ratios}
\end{center}
\end{figure}

In figure \ref{fig:BdNdy_ratios} we show the same results in the
``measured/expected'' form, namely the experimental data divided by
our  calculations. It is tempting to use the difference between this ratio and the unity as a measure of the magnitude of the final state  effect for $J/\psi$ production. However,  at the moment we prefer to refrain from a premature conclusion about the size of the final state effect as such a conclusion would crucially depend on the value of the $C$ factor. In addition we are aware of the limitations due to the accuracy of the experimental data.  To give a firm conclusion we need to perform a  comparison with a high precision $dAu$ data, where the $C$ factor can eventually be fixed with a higher accuracy.

\begin{figure}
\begin{center}
\includegraphics[width=10cm]{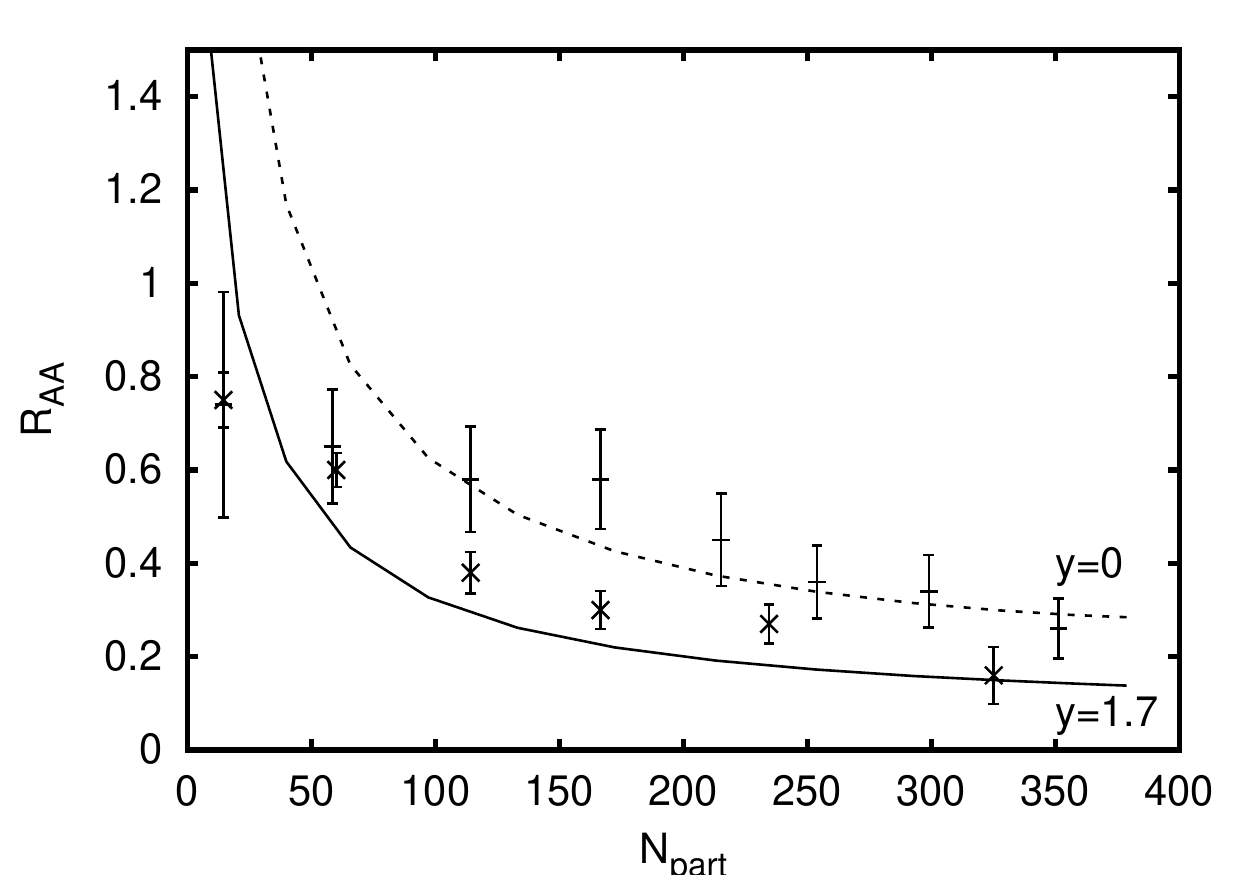}
\caption{Nuclear modification factor for $ J/\psi$ production in heavy ion collisions for different rapidities.}
\label{fig:raa}
\end{center}
\end{figure}

To emphasize the nuclear dependence of the inclusive cross sections it is convenient to introduce the nuclear modification factor
\beq\label{nmf}
R_{AA}(y,N_\mathrm{part})=\frac{ \frac{dN^{AA}}{dy} }
{N_\mathrm{coll}\,\frac{dN^{pp}}{dy}}\,.
\eeq
It is normalized in such a way that no nuclear effect would correspond to $R_{AA}=1$. In \fig{fig:raa} we plot the result of our calculation. The nuclear modification factor exhibits the following  two important features: (i) unlike the open charm production, $ J/\psi$ is suppressed  at $y=0$. This is not very surprising since the probability of the $ J/\psi$ formation is reduced due to multiple interactions with the gluons; (ii) cold nuclear effects account for a significant part of the $ J/\psi$ suppression observed in heavy ion collisions. 
On the other hand, the cold nuclear effects discussed in this paper may not be sufficient to account for a very low $R_{AA}$ in the most central events. 
Higher precision $dA$ and $AA$ data will allow to tell whether there is a suppression of $ J/\psi$ in 
quark--gluon plasma, or the directly produced  $ J/\psi$'s survive \cite{Karsch:2005nk}.

\section{Conclusions}

In this paper we have developed a model for $ J/\psi$ production in heavy ion collisions. Assuming that contributions of strong color fields of the two colliding nuclei do not interfere we summed all the diagrams proportional to the positive powers of the large parameter $\as^2 A^{1/3}\sim 1$ in both nuclei. 
Our main result is given by Eq.~\eq{IP1}. In the RHIC kinematic region this formula can be simplified  and is given by \eq{IP4}. We used this equation in our numerical calculation with the realistic nuclear profiles described in Sec~\ref{sec:numerics} in details. The results are presented in Figs.~\ref{fig:BdNdy}--\ref{fig:raa}. 

We can see that the rapidity and centrality dependence are reproduced quite well. This observation implies  that an appreciable amount of the $ J/\psi$ suppression in high energy heavy ion collisions comes from the cold nuclear effects. Fig.~\ref{fig:raa} demonstrates that the nuclear modification factor for $ J/\psi$ production is \emph{strongly suppressed even at zero temperature.}  While $ J/\psi$ suppression in the forward direction is not a surprise (the nuclear modification factor for light and, perhaps, heavy hadrons is known to be suppressed), similar behavior in the central rapidity region is a peculiar feature of the $ J/\psi$ production. The reason is that multiple scattering of $c\bar c$ pair in the cold nuclear medium increases the relative momentum between the quark and antiquark, which makes the bound state formation less probable. Formally, the sum rule proven in \cite{KKT}, which guarantees emergence of the Cronin enhancement for single partons, is broken for the bound states.

Our result strongly suggests that the cold nuclear matter effects play a very important role in  $ J/\psi$ production in heavy ion collisions. The final nuclear modification factor, which is measured in experiment,  is undoubtedly a result of a delicate interplay between the cold and hot nuclear matter effects.

We consider the present work as the first step towards understanding the role of the cold nuclear effects in $ J/\psi$ production in high energy heavy ion collisions.  We calculated the parametrically enhanced contribution coming from 
even number of scatterings of $c \bar c$ pair in the nuclei. However, we neglected other contributions that may  be phenomenologically important though parametrically small. These include soft gluon radiation in the final state and color octet mechanism of  $ J/\psi$ production.  Moreover, for peripheral collisions these contributions become of the same order as the one discussed in this paper. 
Therefore, we plan to perform a detailed investigation of these contributions in the future.

\vskip0.3cm
{\bf Acknowledgments.}
The work of  D.K. was supported by the U.S. Department of Energy under Contract No. DE-AC02-98CH10886.
K.T. was supported in part by the U.S. Department of Energy under Grant No. DE-FG02-87ER40371; he would like to
thank RIKEN, BNL,
and the U.S. Department of Energy (Contract No. DE-AC02-98CH10886) for providing facilities essential
for the completion of this work.  This research of E.L.  was supported in part by  a
grant from Ministry of Science, Culture \& Sport, Israel \& the
Russian  Foundation for Basic research of  the Russian Federation,
and by BSF grant \# 20004019.
\vskip0.3cm

\appendix
\section{Heavy quark production}\label{sec:quarks}

\begin{figure}[ht]
\begin{center}
   \includegraphics[width=8cm]{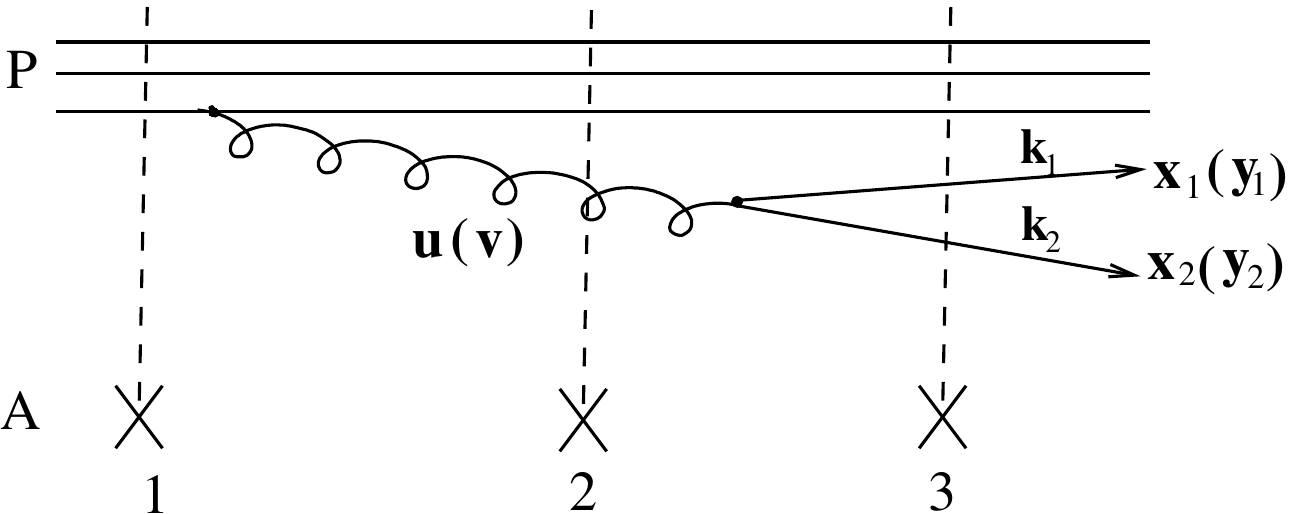} 
 \end{center}    
\caption{Diagrams contributing to the $q\bar q$ production in pA
  collisions at high energies. 
Transverse momenta of the produced quark and antiquark $\un k_1$, $\un k_2$,
their transverse coordinates in the amplitude $\un x_1$, $\un x_2$ and
in the complex conjugated one $\un y_1$, $\un y_2$ and 
the gluon transverse coordinate in the amplitude $\un u$ and in the
complex conjugated one $\un v$ are shown in boldface. 
Vertical dashed lines indicate all possible  interaction times of
incoming proton with the nucleus (shown by crosses). 
 More detailed discussion can be found in
 \cite{Tuchin:2004rb,Kovchegov:2006qn}. Notations follow
 \cite{Kovchegov:2006qn}.} 
\label{fig2}
\end{figure}

In this appendix we derive the cross section for production of a
$c\bar c$ pair with fixed relative momentum in pA collisions. A
general problem of  
 $q\bar q$ pair production in pA collisions including all possible
 nonlinear evolution effects was solved in
 Refs.~\cite{Tuchin:2004rb,Kovchegov:2006qn}.  
 Let us introduce the following notations: $\un k_1$ and $\un k_2$ are
 the produced quark and anti-quark transverse momenta, $\un q$ is the
 gluon transverse momentum, $z=k_+/q_+$ is a fraction of the
 light-cone momentum of gluon carried by the produced quark;  
$\un x_1$ and $\un y_1$ are the transverse coordinates of the produced
quark in the amplitude and in the complex conjugated amplitude
respectively; $\un x_2$ and $\un y_2$ are the corresponding
coordinates of the antiquark. Transverse coordinates of gluon in the
amplitude $\un u$ and in the complex conjugated amplitude $\un v$ are
given by  $\un u \, \equiv \, z \, \un x_1 + (1 - z) \, \un x_2$ with $u =
|\un u|=r$ and $\un x_{12} = \un x_1 - \un x_2\equiv \un r$, ($x_{12} = |\un
x_{12}|$) and analogously for $\un v$. With these notations 
the double inclusive  quark--anti-quark production cross section is
given by \cite{Tuchin:2004rb,Kovchegov:2006qn} 
\beq
\frac{d \, \sigma}{d^2 k_1 \, dk_2dy \, d^2 b} \,=\, \frac{1}{4 \, (2
  \, \pi)^6} \, \int   d^2 x_1 \, d^2 x_2 \, d^2 y_1 \,
d^2y_2\,\int_0^1 dz \, e^{-i \, \un k_1 \cdot (\un x_1-\un 
y_1)-i\un k_2\cdot (\un x_2-y_2)}  \, \eeq
\beq\label{single_cl}
\times\,\sum_{i,j=1}^3\, \Phi_{ij}\, (\un x_1, \un x_2; \un y_1, \un
x_2; z) 
\, \Xi_{ij} (\un x_1, \un x_2; \un y_1, \un x_2; z)\,,
\eeq
The products of the light-cone ``wave functions" are detailed as
follows \cite{Kovchegov:2006qn} 
\ben
\Phi_{11} (\un x_1, \un x_2; \un y_1,\un y_2 ; z) \, = \, 4 \, C_F \, 
\bigg(\frac{\as}{\pi}\bigg)^2 \, \bigg\{ F_2(\un x_1, \un x_2; z) \, 
F_2 (\un y_1, \un y_2; z) \, \frac{1}{x_{12} \, y_{12} \, u \, v} \, 
[ (1 - 2 \, z)^2 
\een
\ben
\times\,  (\un x_{12} \cdot \un u) \, (\un y_{12} \cdot \un v) 
+ (\epsilon_{ij} \, u_i \, x_{12 \, j}) \, (\epsilon_{kl} \, v_k \, y_{12 \, l})] +
F_1(\un x_1, \un x_2; z) \, F_1(\un y_1, \un y_2; z) \, m^2 \, 
\frac{\un u\cdot\un v}{u \, v} 
\een
\ben
+ 4 \, z^2 \, (1-z)^2 \,  F_0(\un x_1, \un x_2; z) \, 
F_0(\un y_1, \un y_2; z)  - 2 \, z \, (1 - z) \, (1 - 2 \, z) \,
\bigg[ \frac{\un x_{12} \cdot \un u}{x_{12} \, u} \, F_2(\un x_1, \un x_2; z) 
\een
\beq\label{phi11}
\times \,
F_0(\un y_1, \un y_2; z) +  
\frac{\un y_{12} \cdot \un v}{y_{12} \, v} \, 
F_2(\un y_1, \un y_2; z) \, F_0(\un x_1, \un x_2; z) \bigg] \, \bigg\} \,,
\eeq
\ben
\Phi_{22} (\un x_1, \un x_2; \un y_1,\un y_2; z) \, = \, 4 \, C_F \,
\bigg(\frac{\as}{\pi}\bigg)^2 \, m^2 \, \bigg\{
K_1(m \, x_{12}) \, K_1(m \, y_{12}) \, \frac{1}{x_{12} \, y_{12} \, u^2 \, v^2} 
\, [ (1 - 2 \, z)^2
\een
\beq\label{phi22}
\times\,(\un x_{12} \cdot \un u) \, (\un y_{12} \cdot \un v) 
+ (\epsilon_{ij} \, u_i \, x_{12 \, j}) \, (\epsilon_{kl} \, v_k \, y_{12 \, l})] +
K_0(m \, x_{12})\, K_0(m \, y_{12}) \, \frac{\un u\cdot\un v}{u^2 \, v^{2}} \bigg\}\,,
\eeq
\ben
\Phi_{12} (\un x_1, \un x_2; \un y_1, \un y_2; z) = -  4 \, C_F \, 
\bigg(\frac{\as}{\pi}\bigg)^2 \, m \,
\bigg\{ F_2 (\un x_1, \un x_2;z) \, K_1(m \, y_{12}) \, 
\frac{1}{x_{12} \, y_{12} \, u \, v^2} \, [ (1 - 2 \, z)^2
\een
\ben
\times\, (\un x_{12} \cdot \un u) \, (\un y_{12} \cdot \un v) 
+ (\epsilon_{ij} \, u_i \, x_{12 \, j}) \, (\epsilon_{kl} \, v_k \, y_{12 \, l})]
+ m \, F_1(\un x_1,\un x_2; z) 
\, K_0(m \, y_{12}) \, \frac{\un u\cdot\un v}{u \, v^{2}} 
\een
\beq\label{phi12}
- 2 \,  z \, (1 - z) \, (1 - 2 \, z) \,
\frac{\un y_{12} \cdot \un v}{y_{12} \, v^2}  \, 
F_0(\un x_1, \un x_2; z) \, K_1 (m \, y_{12}) \bigg\}\,, 
\eeq
\ben
\Phi_{33} (\un x_1, \un x_2; \un y_1, \un y_2; z) \, =\, 
\Phi_{11} (\un x_1, \un x_2; \un y_1, \un y_2; z) + 
\Phi_{22} (\un x_1, \un x_2; \un y_1, \un y_2; z) +
\Phi_{12} (\un x_1, \un x_2; \un y_1, \un y_2; z) 
\een
\beq\label{phi33}
+ \Phi_{21} (\un x_1, \un x_2; \un y_1, \un y_2; z)
\eeq
\beq\label{phi13}
\Phi_{13} (\un x_1, \un x_2; \un y_1, \un y_2; z) \, =\, 
- \Phi_{11} (\un x_1, \un x_2; \un y_1, \un y_2; z) 
- \Phi_{12} (\un x_1, \un x_2; \un y_1, \un y_2; z)
\eeq
\beq\label{phi23}
\Phi_{23} (\un x_1, \un x_2; \un y_1, \un y_2; z) \, =\, 
- \Phi_{21} (\un x_1, \un x_2; \un y_1, \un y_2; z) 
- \Phi_{22} (\un x_1, \un x_2; \un y_1, \un y_2; z)
\eeq
\beq\label{phi_symm}
\Phi_{ij} (\un x_1, \un x_2; \un y_1, \un y_2; z) \, =\, 
\Phi_{ji}^* (\un y_1, \un y_2; \un x_1, \un x_2; z).
\eeq
The auxiliary functions $F_1$, $F_2$ and $F_0$ are defined as 
\begin{eqnarray}
F_2 (\un x_1, \un x_2; z)&=& \int_0^\infty dq \, J_1 (q \, u) \,
K_1 \bigg(x_{12} \, \sqrt{m^2+
q^2\, z(1-z)} \bigg)\, \sqrt{m^2+
q^2\, z(1-z)} \,,\label{aux.fun1}\\ 
F_1 (\un x_1, \un x_2;z)&=& \int_0^\infty dq \, J_1(q \, u) \, K_0 \bigg(x_{12}
\, \sqrt{m^2+q^2\, z(1-z)} \bigg)\,,\label{aux.fun2}\\
F_0 (\un x_1, \un x_2;z)&=& \int_0^\infty dq \, q \, J_0(q \, u) \, K_0 \bigg(x_{12}
\, \sqrt{m^2+q^2\, z(1-z)} \bigg)\,,\label{aux.fun3}
\end{eqnarray}
where $u = |\un u|$, $\un x_{12} = \un x_1 - \un x_2$, $x_{12} =
|\un x_{12}|$, and $\un q = \un k_1 + \un k_2$. 

By definition, the \emph{gluon} saturation scale $Q_s$ 
\beq\label{gluon.sat}
Q_s^2 \, = \, 4 \, \pi \, \as^2  \, \rho \, T(\un b)
\eeq
with $\rho$ the nucleon number density in the nucleus and $T(\un b)$
the nuclear profile function. We also use the \emph{quark} saturation
scale $\mathcal{Q}^2_s$  give by  
\beq\label{quark.sat}
\mathcal{Q}^2_s\,=\, \frac{C_F}{N_c}\,Q_s^2\stackrel{N\to \infty}{\longrightarrow}
\frac{1}{2}\, Q_s^2\,.
\eeq
Throughout the theoretical discussion we assumed  for simplicity that
the nuclear profile is cylindrical with $T(\un b)\approx 2R_A$ for
$\un b^2\le R_A^2$. However, numerical calculations in
Sec.~\ref{sec:numerics} are performed with an accurate
parameterization as discussed there in detail. In the large $N_c$
approximation we write 
\begin{eqnarray}
\Xi_{11} (\un x_1, \un x_2; \un y_1, \un y_2; z) 
&=&e^{-\frac{1}{8}\, (\un x_1 -\un y_1)^2 \, Q_s^2 \, \ln 
(1/|\un x_1 -\un y_1| \, \mu)
-\frac{1}{8}\, (\un x_2 -\un y_2)^2 \, Q_s^2 \, 
\ln (1/|\un x_2 -\un y_2| \, \mu)}\,,\label{xi1}\\
\Xi_{22} (\un x_1, \un x_2; \un y_1, \un y_2; z) 
&=&e^{-\frac{1}{4}\, (\un u-\un v)^2 \, Q_s^2 \, 
\ln (1/ |\un u -\un v| \, \mu)}\,,\label{xi2}\\
\Xi_{33} (\un x_1, \un x_2; \un y_1, \un y_2; z) &=& 1\,,\label{xi3}\\
\Xi_{12} (\un x_1, \un x_2; \un y_1, \un y_2; z) &=& e^{-\frac{1}{8}\, 
(\un x_1 -\un v)^2 \, Q_s^2 \, 
\ln (1/ |\un x_1 -\un v| \, \mu) -\frac{1}{8}\, (\un x_2 -\un v)^2 \, Q_s^2 \, 
\ln (1/ |\un x_2 -\un v| \, \mu)}\, ,\label{xi4}\\
\Xi_{23} (\un x_1, \un x_2; \un y_1, \un y_2; z) &=& e^{-\frac{1}{4} \, u^2 \, 
Q_s^2 \, \ln (1/u \, \mu)}\,,\label{xi5}\\
\Xi_{13} (\un x_1, \un x_2; \un y_1, \un y_2; z) &=& e^{-\frac{1}{8} \, x_1^2 \, 
Q_s^2 \, \ln (1/x_1 \mu) - \frac{1}{8} \, x_2^2 \, Q_s^2 \, \ln (1/x_2
\mu)}\,\label{xi6} 
\end{eqnarray}
 All other $\Xi_{ij}$'s can be found from the components listed in
\eq{xi1}-\eq{xi6} using
\beq
\Xi_{ij} (\un x_1, \un x_2; \un y_1, \un y_2; z) \, = \, 
\Xi_{ji} (\un y_1, \un y_2; \un x_1, \un x_2; z)
\eeq
similar to \eq{phi_symm}. 

If the typical gluon momentum $\un q$ is much smaller than the
produced quark mass, the above expressions can be  significantly
simplified. Indeed, since $z(1-z)\leq 1/4$ we get 
 \beq\label{approx}
  \un q^2z(1-z)\ll m^2\,,
  \eeq
   and the auxiliary functions read
\begin{eqnarray}\label{aux.simpl}
F_2(\un x_1,\un x_2;z)&=&K_1(x_{12}m)\,m\,u^{-1}\,,\\
F_1(\un x_1,\un x_2;z)&=&K_0(x_{12}m)\,u^{-1}\,,\\
F_0(\un x_1,\un x_2;z)&=&0\,.
\end{eqnarray}
In this approximation the only non-vanishing products of  ``wave functions" are given by
$$
\Phi_{11}(\un x_1, \un x_2; \un y_1, \un y_2; z) \, =\, 
\Phi_{22}(\un x_1, \un x_2; \un y_1, \un y_2; z) \, =\, -
\Phi_{12}(\un x_1, \un x_2; \un y_1, \un y_2; z) \, 
$$
\ben
 = \, 4 \, C_F \, 
\bigg(\frac{\as}{\pi}\bigg)^2 \,m^2\, \bigg\{ K_1(x_{12}m)\,
K_1(y_{12}m)\,\frac{1}{x_{12} \, y_{12} \, u^2 \, v^2} \,  
[ (1 - 2 \, z)^2 \,(\un x_{12} \cdot \un u) \, (\un y_{12} \cdot \un v) 
\een
\ben
+ (\epsilon_{ij} \, u_i \, x_{12 \, j}) \, (\epsilon_{kl} \, v_k \, y_{12 \, l})] +
K_0(x_{12}m)\,K_0(y_{12}m)\,\frac{\un u\cdot\un v}{u^2 \, v^2} \bigg\}\,.
\een
Averaging over all directions of  gluon emission from the valence quark using 
$\langle \epsilon_{ij} \, u_i \, x_{12 \, j} \, \epsilon_{kl} \, v_k \, y_{12 \, l}\rangle = 
u\,v\,\un x_{12}\cdot \un y_{12}$ we arrive at the well-known result
\cite{Tuchin:2004rb,KopTar} 
$$
\Phi_{11}(\un x_1, \un x_2; \un y_1, \un y_2; z) = 
$$
\beq\label{phiT}
4 \, C_F \, 
\bigg(\frac{\as}{\pi}\bigg)^2 \,\frac{m^2}{uv}\,\bigg\{ \frac{\un
  x_{12}\cdot \un
  y_{12}}{x_{12}y_{12}}[(1-z)^2+z^2]\,K_1(x_{12}m)\,
K_1(y_{12}m)+K_0(x_{12}m)K_0(y_{12}m)\bigg\} 
\eeq

Let's now introduce the following notations, see \fig{ccpp}:
\beq
\un p=\un k_1+\un k_2\,,\quad \un k= \un k_1-\un k_2+(1-2z)\, (\un k_1+\un k_2)\,,
\eeq
or, equivalently,
\beq
\un k_1=z\un p+\frac{1}{2}\un k\,,\quad \un k_2=(1-z)\,\un p-\frac{1}{2}\,\un k\,.
\eeq
At $z=1/2$, momentum $\un k$ becomes the relative momentum of the
$c\bar c$ pair. In this case integration over the total momentum $\un
p$ results in the delta function  
$\delta^{(2)}(\un u-\un v)$. This delta-function simplifies
expressions in the exponents of \eq{xi1}--\eq{xi6}.     
For the sum over all rescattering factors including the signs of
$\Phi_{ij}$'s we get (omitting the logarithms  $\ln(1/x\mu)$ for
brevity) 
\beq\label{sumX}
\Xi(\un x_{12},\un y_{12};z=1/2)=e^{-\frac{1}{8}(\un x_{12}-\un y_{12})^2 \, Q_s^2}+1-
e^{-\frac{1}{8}\un x_{12}^2Q_s^2}-e^{-\frac{1}{8}Q_s^2y_{12}^2}
\eeq

Using the delta function in \eq{single_cl} to integrate over $\un u$
and integrating over $\un v$ in the leading logarithmic approximation
we derive  
$$
\frac{d\sigma}{d^2k \,dy\, d^2b}=\frac{C_F\,\as^2\,m^2}{4 \pi^5}\int_0^1 dz
\int d^2x_{12}\int d^2y_{12}\, e^{-i\frac{1}{2} \un k\cdot(\un
  x_{12}-\un y_{12})}\,\ln(1/\mu|\un x_{12}-\un y_{12}|) 
$$
\beq\label{main0}
\times \bigg\{\frac{\un x_{12}\cdot \un
  y_{12}}{x_{12}y_{12}}[(1-z)^2+z^2]\,K_1(x_{12}m)\,
K_1(y_{12}m)+K_0(x_{12}m)K_0(y_{12}m)\bigg\}\, \Xi(\un x_{12},\un
y_{12};z) 
\eeq
Finally, using the $\un x_{12}=\un r$, $y_{12}=\un r'$ notation obtain
the final expression for the $c\bar c$ pair production with fixed
relative momentum (at $z=1/2$): 
$$
\frac{d\sigma_{pA}}{d^2k \,dy\, d^2b}=\frac{\as\,m^2}{8 \pi^4}
\int d^2r\int d^2r'\, e^{-i\frac{1}{2} \un k\cdot(\un r-\un
  r')}\,xG(x_1,m_c^2)\bigg\{\frac{1}{2} \frac{\un r\cdot \un
  r'}{rr'}\,K_1(rm)\, K_1(r'm)+K_0(rm)K_0(r'm)\bigg\}\, 
$$
\beq\label{main}
\times \bigg\{ e^{-\frac{1}{8}(\un r-\un r')^2 \, Q_s^2}+1-
e^{-\frac{1}{8}\un r^2Q_s^2}-e^{-\frac{1}{8}Q_s^2r'^2}\bigg\}\, 
\eeq

To generalize this expression in the case of heavy ion collisions we
 replace  the
proton's gluon distribution function  by the correlation function of
the color field potentials of the second nucleus as follows  \cite{Kovchegov:2000hz} 
\beq\label{replace}
\frac{d\,xG(x,4/r)}{d^2b}= 
\frac{4\,N_c}{\pi^2\as\, r^2}\, \left(1-e^{-\frac{1}{4}r^2\Q_s^2}\right)\,.
\eeq
Therefore, taking the large $N_c$ approximation $\Q_s^2\approx
\frac{1}{2}Q^2$ and $C_F\approx N_c/2$, we derive 
$$
\frac{dN_{AA}}{d^2k \,dy\, d^2b}=\frac{\as\,m^2}{8 \pi^4}
\int d^2r\int d^2r'\, e^{-i\frac{1}{2} \un k\cdot(\un r-\un
  r')}\,\bigg\{\frac{1}{2} \frac{\un r\cdot \un r'}{rr'}\,K_1(rm)\,
K_1(r'm)+K_0(rm)K_0(r'm)\bigg\}\, 
$$
$$
\times \frac{8\,C_F}{\pi^2\as}\,
 \bigg\{ \frac{1}{r^2}\left(1-   e^{-\frac{1}{8}\un
     r^2Q_{s1}^2}\right) \left(1-   e^{-\frac{1}{8}\un
     r^2Q_{s2}^2}\right)+  
\frac{1}{r'^2} \left(1-   e^{-\frac{1}{8}\un r'^2Q_{s1}^2}\right)
\left(1-   e^{-\frac{1}{8}\un r'^2Q_{s2}^2}\right)\,. 
$$
\beq\label{main2}
   -   
\frac{1}{(\un r-\un r')^2}\left(1-   e^{-\frac{1}{8}(\un r-\un
    r')^2Q_{s1}^2}\right) \left(1-   e^{-\frac{1}{8}(\un r-\un
    r')^2Q_{s2}^2}\right)    
   \bigg\}\, .
   \eeq


\end{document}